\documentclass[a4paper,fleqn]{cas-dc}

\usepackage[authoryear]{natbib}

\usepackage{gensymb}

\usepackage[switch, modulo]{lineno}
\nolinenumbers

\usepackage[dvipsnames]{xcolor}

\newcommand\fix[1]{\textcolor{Red}{#1}}

\begin{document}

\let\WriteBookmarks\relax
\def\floatpagepagefraction{1}
\def\textpagefraction{.001}

\shorttitle{5G on the Farm}

\shortauthors{T Zhivkov et~al.}

\title [mode = title]{5G on the Farm: Evaluating Wireless Network Capabilities for Agricultural Robotics}                      
\tnotemark[1]

\tnotetext[1]{This document is the results of the research project funded by UKRI Research England under the Lincoln Agri-Robotics and by Ceres Agri-tech grants.}


\author[1]{Tsvetan Zhivkov}[type=editor,
                        auid=000,bioid=1,
                        prefix=,
                        role=Postdoctoral Researcher,
                        orcid=XXXX-XXXX-XXXX-XXXX]

\cormark[1]

\fnmark[1]

\ead{tzhivkov@lincoln.ac.uk}



\author[1]{Elizabeth I Sklar}[type=editor,
                        auid=000,bioid=1,
                        prefix=,
                        role=Professor,
                        orcid=0000-0002-6383-9407]

\ead{esklar@lincoln.ac.uk}

\affiliation[1]{organization={Lincoln Institute for Agri-food Technology, University of Lincoln},
    addressline={Riseholme Park Campus},
    city={Lincoln},
    postcode={LN2 2LG}, 
    country={UK}
    }

\cortext[cor1]{Corresponding author}



\begin{abstract}
\fix{%
Global food security is an issue that is fast becoming a critical matter in the world today.
Global warming, climate change and a range of other impacts caused by humans,
such as carbon emissions, sociopolitical and economical challenges (e.g. war), traditional workforce/labour decline and population growth are straining global food security.
The need for high-speed and reliable wireless communication in agriculture is becoming more of a necessity rather than a technological demonstration or showing superiority in the field.
Governments and industries around the world are seeing more urgency in establishing communication infrastructure to scale up agricultural activities and improve sustainability, by employing autonomous agri-robotics and agri-technologies.
The work presented here evaluates the physical performance of 5G in an agri-robotics application, and the results are compared against 4G and WiFi6 (a newly emerging wireless communication standard), which are typically used in agricultural environments.
In addition, a series of simulation experiments were performed to assess the ``real-time'' operational delay in critical tasks that may require a human-in-the-loop to support decision making.
The results lead to the conclusion that 4G cannot be used in the agricultural domain for applications that require high throughput and reliable communication between robot and user.
Moreover, a single wireless solution does not exist for the agricultural domain, but instead multiple solutions can be combined to meet the necessary telecommunications requirements.
Finally, the results show that 5G greatly outperforms 4G in all performance metrics, and on average only 18.2ms slower than WiFi6 making it very reliable.
}
\end{abstract}



\maketitle


\section{Introduction}
\label{sec:introduction}

The agricultural domain is currently experiencing increased focus on emerging technologies and robotic applications, largely due to concerns about food security and responses to climate change. 
As a result, the intersection of agricultural robotics (or \emph{agri-robotics}) and supporting technologies, like telecommunications,
has garnered international attention from governments and industries alike~\citep{gov:fund,gov:treeplant-report,duckett:agri-robots,zhivkov:5g,tang-survey:5g}.

Some of the key reasons behind the recent boom in agricultural robotics include increased negative impact on global food security derived from a range of sociopolitical and environmental factors.
The current war between Russia and Ukraine is one example where sociopolitical events have stressed the food supply chain.
Ukraine and Russia are estimated to account for 30\% of the global wheat supply~\citep{Lango:ukraine-wheat,rae:rus-ukraine-war}.
Another example is Brexit, where the UK leaving the single market has led to fewer seasonal workers entering the UK, traditionally from Eastern Europe, to support harvesting of fruits and vegetables~\citep{partridge-partington:2021}.
The Covid-19 pandemic laid bare the impact of labour shortages on agricultural productivity in multiple countries~\citep{washburn:2020,naik:2020}.
Examples of environmental effects on agriculture go both ways.
Climate change affects agriculture by causing unpredictable changes in patterns of rainfall, average temperatures, occurrence of heatwaves, prevalence of weeds, infestations of pests, etc., all of which can stress and damage crops~\citep{raza:climate-change,webb:growingpop-climatechange}.
Another issue related to climate change that is producing significant negative impact on agriculture  is soil erosion.
While some amount of soil erosion occurs naturally because of environmental factors, it is more unpredictable and more likely to occur due to climate change~\citep{borrelli:soil-climate}.
On the flip side, agriculture affects the environment negatively in multiple ways.
For example, current intense farming practices cause soil compaction~\citep{shaheb:soil,millard:soil-impact} due to heavy farm vehicles continuously driving over farm fields.
According to the UN's Food and Agriculture Organisation, livestock account for significant percentage of the greenhouse gas (GHG) emissions caused by humans~\citep{fao-gleam:2022}.

One of the primary motivators for the research initiative on \textit{sustainable farming} is population growth~\citep{webb:growingpop-climatechange}: current agricultural practises are not sustainable and are not viable for up-scaling to secure food for the predicted 10.2 billion people by 2050~\citep{borsellino:food-sustainability}.
\emph{Precision agriculture}~\citep{stafford:2000,gebbers-adamchuk-science:2010} encompasses a broad spectrum of intelligent technologies that allow growers to make decisions at the level of an individual plant, or group of contiguous plants, rather than an entire field. This means that resources, such as fertilisers and water, as well as herbicides and pesticides, can be targeted to specific plants or planted regions, rather than applying across an entire field---which can be wasteful when only a portion of the field requires that treatment, as well as unnecessarily expensive and potentially damaging to the environment.
Intelligent sensors, either mounted in static locations around fields or on mobile devices such as ground-based or aerial (UAV) robots, can feed precise location-specific information about plant growth (e.g. size, colour, shape) and the environment (e.g. temperature, moisture, humidity) to farmers, who can use that data to inform their timetables concerning when (or not) to spray, when to harvest, etc.
In addition, specialised actuators can also be mounted on robots to allow the location-specific information to feed real-time decision making and trigger actuation such as spraying, mechanical weeding or harvesting, performed by robots in the field.

From the \emph{intelligent robotics} perspective, these types of tasks require a number of capabilities:
(a) precise location of sensed information;
(b) precise location for actuation;
(c) path-planning for ground-based robots to minimise soil compaction and energy usage or flight planning for aerial robots to minimise energy usage;
and
(d) accurate analysis of sensed data.
The joint desire for highly accurate position information and the ability to transmit sensor data from fields to farmers (or automated decision support systems) highlights the need for communications networks that can deliver both of these capabilities reliably and robustly.
5G telecommunications is anticipated to provide three key technical advantages over existing 4G capabilities~\citep{durisi-et-al:2016}: \emph{enhanced mobile broadband (eMBB)} at high bandwidth, \emph{ultra reliable low-latency (URLLC)} and \emph{massive machine communications (mMTC)} at low bandwidth high scale and within a sliced network architecture~\citep{popovski20185g}.
Taken together, eMBB and URLLC promise better performance for applications such as accurate and timely positioning and faster sensor data transmission.
However, 5G is not available everywhere and public installations, especially in rural communities where populations are sparse (i.e. the paying customer base is small), are not a high priority in many countries.
Thus we are concerned with understanding the specific practical advantages of 5G within an agriculture application domain.

In the work presented here, we explore three wireless network technologies (WiFi6, 5G and 4G) and 
evaluate their performance in an agri-robotics environment.
More emphasis is placed on 5G and WiFi6, as these are new and emerging communication networks, whereas we consider 4G as our baseline.
Two separate experiments are conducted, a physical experiment and a simulation experiment.
The physical experiment
tests the application of two-way communication between an in-field robot streaming real-time video to a remote server, which performs detection of crops and weeds from the received video.
The received video stream is converted to a stream of images that are passed on to an AI-driven detection system.
The wireless network results for throughput and latency are compared.
The simulation experiment uses the average latency results from the physical experiment 
to compare ``real-time'' operational performance.

The remainder of this paper is comprised of
a literature review (Section~\ref{sec:litreview}), exploring the current research and state-of-the-art; 
our methodology and testing environment (Section~\ref{sec:methodology}), discussing the three wireless networks and locations where our experiments were conducted;
our physical experiment (Section~\ref{sec:physical:exp}), analysing network throughput and latency;
our simulation experiment (Section~\ref{sec:sim:exp}), investigating the performance of ``real-time'' operation and control;
and finally a conclusion (Section~\ref{sec:conclusion}).

\section{Related Work}
\label{sec:litreview}

Agricultural environments hold many challenges for wireless networks, as they are unstructured and have natural obstacles that cannot be penetrated by most forms of telecommunications.
Alternatively, laying wired infrastructure around farms is even more challenging and expensive~\citep{schneir:rural-telecoms}.
Because of the shear sizes and remote locations of farms, it is unrealistic to dig trenches or erect overhead structures and permanently place wiring or optical cables to provide communication across fields.
In addition, farmers are prone to dig up land occasionally, which might offset or damage underground cables, as well as introduce pests (e.g. termites); and finally, heavy vehicles can cause soil compaction, especially in wet conditions, between 25 cm and 50 cm deep, which can move or damage cables.
Fibre optic cables are usually placed 15 cm to 20 cm underground, but can be laid much deeper with a significant increase in cost.
Instead, smaller 5G cells can be used to deliver high-speed and reliable wireless communication to rural areas~\citep{tang-survey:5g}.
Of course, the cost of 5G carrier and user equipment remains very high and such a 5G expanse is yet to arrive in rural areas.

From all the challenges that come with wired infrastructure, it is no wonder that the literature in communication and networking for agriculture is mainly focused on wireless networks, Internet-of-Things (IoT), low-cost and low-power sensors~\citep{adami:agri-IoT,kagan:IoT-review,Tao:IoT-review}.
The research looks at either 
multiple sensors on a single device (System-on-Chip), performing a specific function,
or cloud/fog computing for data collection~\citep{Karthikkumar:Iot,Bodunde:IoT,Tsipis:Iot-cloud}.
WiFi~\citep{Karthikkumar:Iot} and Hybrid WiFi/Zigbee~\citep{Tsipis:Iot-cloud}-based communication infrastructure has been investigated, but is heavily limited in functionality and device support, as well as number of communicating devices.
Little consideration is given to a more practical and permanent communication infrastructure supporting a wide range of devices, functionalities and robot operations in agricultural environments.
However, 5G has not only the potential to support a large number of connected devices, but also a wide range of different devices~\citep{zhivkov:5g}.
Moreover, considering the growth and popularity of 3G and 4G as use cases, the number of devices that support 5G will continue to grow as the technology matures, bringing cheaper and lower power devices to the market.
The lack of such research has been noted with the emergence of 5G~\citep{tang-survey:5g,maraveas:iot-5g}.

The literature in robotics for agriculture shows that the right research questions are being investigated, however progress is much slower than expected, especially for agri-robotics.
A divide can be seen between emerging robot applications in other fields such as warehouse robotics, human-robot interaction (HRI) and self-driving autonomous vehicles, which employ state-of-the-art research in planning, navigation, environment interaction and machine vision.
Comparatively, research in the agri-robotics field is progressing slower; for example, current high-end industrial tractors with RTK-GPS (Real-Time Kinematic positioning GPS) with basic autopilot and mission planner software features~\citep{batte:rtk-tractor} can support much of the emerging research that is being shown off in the agri-robotics field~\citep{panfilov:harper-adams}.
Agri-robotics is lagging behind other fields in incorporating up-and-coming novel deep-learning and machine-learning techniques, as well as not utilising bleeding-edge planning, navigation and machine vision systems to enable completely autonomous operation.

The increased interest in 5G communication technology,
in parallel with the negative sociopolitical and environmental challenges mentioned earlier (e.g. war and Brexit) that have increased government funding in agriculture, have started to shift the balance of agri-robotics and agri-technology onto the right track.
Recent research in agriculture is making gains towards the bleeding edge, 
incorporating novel image detection and machine learning in agri-robotics applications~\citep{gomez:deep-regression,taeyeong:ssupervised-agri}, 
performing novel fleet-management and navigation in agri-robots~\citep{das:agrirobot-navigation} and 
employing 5G networks~\citep{zhivkov:5g}.
These activities motivate and accelerate the need for superior telecommunications infrastructure and innovation in rural environments.

\cite{tang-survey:5g} present use cases and similar research to draw a hypothetical argument for the use of 5G in agriculture and review outcomes based on expectations.
But no actual results are shown of 5G-SA (Stand Alone) in an agricultural application, which is the motivation of the work we present here.
To the best of our knowledge, no research exists in the agri-domain that evaluates and discusses the practical performance of 5G and compares general wireless technology in a rural environment with detailed performance metrics.
Here, we present practical results from a physical experiment performed in two different fields, detailed in Section~\ref{sec:physical:exp}.
In addition, simulated results are drawn from the collected data to further demonstrate the power of 5G in a rural environment, described in Section~\ref{sec:sim:exp}.

\section{Experiment Design}
\label{sec:methodology}

This section describes 
the underpinning agri-robotics use case that serves as the basis for the experiments presented here (Section~\ref{sec:method:application}),
the image detection methodology we implemented for weed identification (Section~\ref{sec:method:ML}),
the locations where experiments were conducted (Section~\ref{sec:method:locations}),
the apparatus deployed (Section~\ref{sec:method:app}),
the wireless network systems evaluated (Section~\ref{sec:method:wireless}),
and 
finally how \emph{tunnelling} was used to support 2-way 5G communication (Section~\ref{sec:method:tunnel}).

\subsection{Agri-robotics use case}
\label{sec:method:application}
The agri-robotics use case we employ for the experiments presented here has two overarching goals: \\
(1)~to develop a sprayer robot that can autonomously drive in a farm field performing real-time weed detection and spot spraying~\citep{salazar-gomez-et-al-iros:2022};
and \\
(2)~to stream real-time video of detected weeds in order to support decision making with a human (e.g. farmer or agronomist) in the loop.

The experimental results and discussion in Section \ref{sec:physical:exp}, evaluate the performance of three wireless network technologies to support goal~(1).
In addition, a proof-of-concept theoretical experiment is conducted in Section~\ref{sec:sim:exp}, in support of the real-time data transmission requirement for achieving goals~(1) and~(2).
The theoretical experiment, Section \ref{sec:sim:exp}, uses latency data obtained from Section \ref{sec:physical:exp}.
The basic setup can also be used for a range of other applications---not only spot spraying for weeds, but also pesticides, spot irrigation and for harvesting.


\begin{figure}[ht]
\begin{center}
\includegraphics[width=0.9\columnwidth]{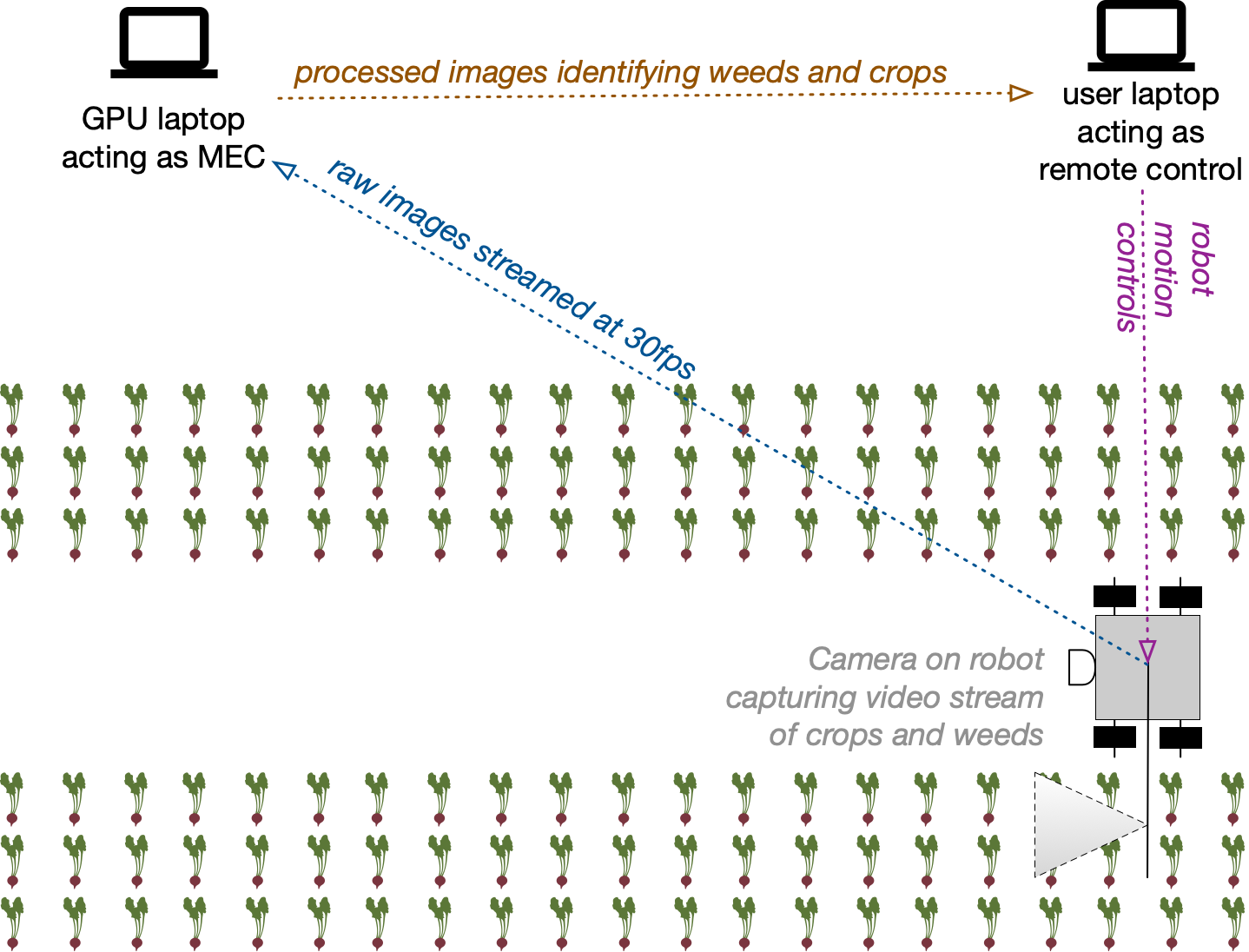}
\caption{Agricultural use case: detecting weeds and crops in a field}
\label{fig:agri-setup}
\end{center}
\end{figure}

\begin{figure}[ht]
\begin{center}
\includegraphics[width=0.9\columnwidth]{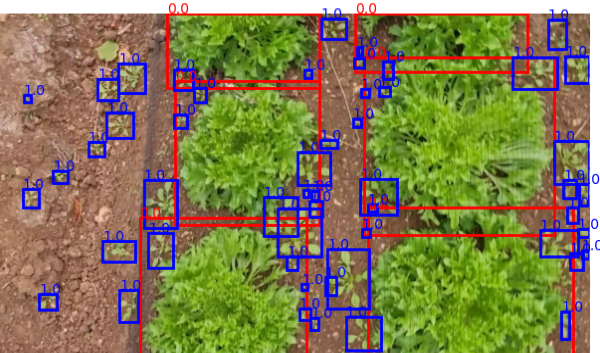}
\caption{Sample detection results of weeds and crop. The red boxes indicate crop (lettuce) and the blue boxes indicate weeds.
A trained Yolov5m model was used to obtain these results, as described in Section~\ref{sec:method:ML}.
}
\label{fig:detection-results}
\end{center}
\end{figure}

The setup of the system is shown in Figure~\ref{fig:agri-setup}.
Wireless communication was established between a remote-controlled robot that was driven in a field, streaming video to a remote laptop with a dedicated GPU acting as a pseudo \emph{Mobile Edge Computer (MEC)} (see Section~\ref{sec:method:app}), which in turn performed image analysis 
in real-time and displayed the results on a screen.
Sample results are shown in Figure~\ref{fig:detection-results}.
The remote laptop was placed at a fixed location, depending on the type of wireless network being tested, as described in Section~\ref{sec:method:wireless}.
As mentioned, the focus of this experiment was to evaluate if real-time weed detection was possible, which would allow a farmer to visualise the performance of the detector and help with their decision making.
We achieve this by evaluating the performance of three different wireless communication networks, as explained below.

\subsection{Image Detection}
\label{sec:method:ML}
In related work~\citep{salazar-gomez-et-al-iros:2022}, we developed a \emph{Machine Learning (ML)} model designed to meet specific image resolution, processing speed and detection accuracy requirements:
\begin{itemize}
    \item To detect weeds accurately using this model, images must be in focus and with a resolution of at least $640 \times 360$ pixels ($\mathit{width} \times \mathit{height}$).
    The ML model benefits from higher resolution images as more detail is retained. 
    \item To achieve ``real-time'' performance, the image processing pipeline (including image capture and object detection) must be capable of running faster than a video stream of 30 \emph{Frames Per Second (FPS)} or higher, which is \textasciitilde33.3ms.
    This is to enable video footage to run uninterrupted at 30FPS with overhead for missed frames.
    \item To provide practical utility for the spot-spraying task at hand, the model needs to achieve >80\% accuracy in crop vs weed detection.
\end{itemize}
In that work, seven different ML models were compared and out of those Yolov5m~\citep{yolov5} achieved the best results out of all requirements.
Yolov5m achieved an accuracy of over 87\% and could perform image inference at a rate of \textasciitilde69FPS~\citep{salazar-gomez-et-al-iros:2022}.

The Yolov5m model was set up on the MEC and the remote-controlled robot streamed video images of the field at 30FPS to the MEC.
The MEC then analysed the incoming video stream, performed image inference using the learned model, created bounding boxes outlining the crops and weeds in each image, and finally displayed a live video feed with the detected weeds and crops to the user.
Figure~\ref{fig:detection-results} illustrates sample results running this model on images of lettuce and surrounding weeds in the field.


\subsection{Experiment Locations}
\label{sec:method:locations}

Experiments were conducted in two fields, 
which we refer to as the \textit{Vegetable Polytunnel} and the \textit{Walled Garden}.
Our 5G network has a geographical advantage in the Vegetable Polytunnel compared to the Walled Garden.
The Vegetable Polytunnel area has \emph{VLoS (Visual Line-of-Sight)} with few obstacles blocking the signal and is at an approximate distance of 46 to 80 metres to the antenna, the closest and furthest points measured respectively.
In contrast, the Walled Garden is important in testing the limitations of the 5G network because it contains regions with \textit{NVLoS} (\textit{No Visual Line-of-Sight}), which are either lightly or heavily obscured by a high tree line and a wall surrounding the field.
Data collection points in the Walled Garden are at approximate distances of 122 to 154 metres, the closest and furthest points measured respectively.
The two areas used for experiments are illustrated in Figures~\ref{fig:newpoly} and~\ref{fig:walledgarden}, in Section \ref{sec:physical:exp}.
The areas of operation and exact distances between data collection points, i.e., network access point (pseudo-MEC) to remote-controlled robot, are given in Section \ref{sec:physical:exp}.

\subsection{Apparatus}
\label{sec:method:app}

The setup of each of the three communications networks compared in this paper are detailed here.
Our 5G system is a stand-alone (SA) network, using the emerging New Radio (NR) sub-6GHz band N77, that is privately owned by our research facility,
making it easier to conduct controlled experiments and with fewer restrictions than a public network.
We are able to adjust certain system parameters, within the constraints of our license agreement, in order to support different types of experimentation.
WiFi, and by extension WiFi6, can be set up as either a private or public network, as it is not controlled by a regulatory body that requires a license to operate\footnote{WiFi6 frequency ranges marginally differ between countries.}.
In the experiments reported here, WiFi6 with the 802.11ax standard was deployed and set up as a private network.
The 4G network in these experiments is commonly used: a commercial, publicly available telecommunications system, with no parameters controlled by end users.
The wireless networks' configuration details and common parameters are discussed further in Section~\ref{sec:method:wireless}.
 
Our 5G-SA system currently does not have a permanent Mobile Edge Compute (MEC) node installed\footnote{Supply-chain issues have delayed acquisition and deployment of all components for the full system, due for completion in 2023.}, which is typically a powerful server-grade system that is used to perform fast computation on the ``edge'' (i.e. in the local environment) as opposed to sending data off to the ``cloud''.
In our setup, the server-grade MEC functionality is approximated by a temporary solution, a powerful GPU-driven laptop, which we refer to as our \textit{pseudo-MEC}.

All experiments were conducted using two laptops that have identical hardware and adequate compute power.
The laptops are deemed to have ``adequate'' processing power if they have a dedicated GPU with at least 4GB or more of graphics RAM.
Each laptop is an ASUS TUF Dash F15\footnote{\url{https://www.asus.com/uk/Laptops/For-Gaming/TUF-Gaming/}}, with i7 11370H @4.8GHz (4 core, 8 thread) CPU, RTX 3060 GPU with 6GB GDDR6 and 8GB DDR4 RAM.
One laptop was used as the remote server,
denoted as the \textit{pseudo-MEC}, and the other acted as a mobile client integrated on a remote-controlled robot in the field.

\subsection{Wireless Networks}
\label{sec:method:wireless}
The network equipment, including the two laptops used for communication experiments, had different setups depending on the type of network being tested.
We tried to keep the setups as similar as possible so that our comparisons of experimental results are valid.
This section describes our three different network setups and configurations.

\begin{figure}[thpb]
\begin{center}
\includegraphics[width=0.9\columnwidth]{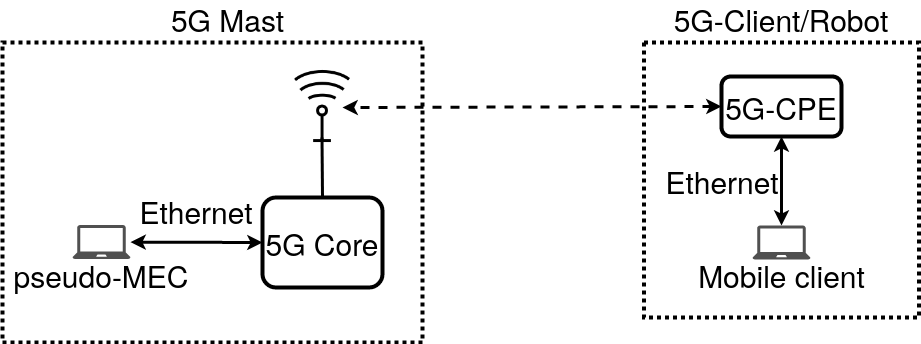}
\caption{Connection diagram for our 5G-SA network.}
\label{fig:5g-clientdiagram}
\end{center}
\end{figure}

\begin{itemize}
\item \textbf{5G Network}: 
A connection diagram is illustrated in Figure~\ref{fig:5g-clientdiagram}.
The pseudo-MEC, for the 5G network experiments is directly attached via Ethernet cable (cat6) to the receiving 5G mast.
Moreover, all Ethernet wired cable connections use cat6 cabling, unless otherwise specified.
The mobile client is connected via an external 5G \emph{CPE (Customer Premises Equipment)} device, by Ethernet cable, to allow it to communicate with the 5G network.
The 5G-CPE is a router using a pre-configured 5G SIM card, provided by BT\footnote{\url{https://www.bt.com/}} and Nokia\footnote{\url{https://www.nokia.com/}}.

\paragraph{5G Network (N77) Configuration:}
The private 5G network system and
relevant parameters are listed in Table~\ref{tab:5gsystem}.
There are certain configuration limitations with our 5G system; the configuration listed in Table~\ref{tab:5gsystem} illustrates what it is capable of achieving at the moment. 
For example, currently the Time Division Duplex (TDD) and carrier bandwidth is fixed, which itself is subject to Ofcom\footnote{\url{https://www.ofcom.org.uk/home}} licensing limitations.
\textsf{DL} and \textsf{UL} stand for \textit{download} and \textit{upload}, respectively, and are used typically to denote throughput speed or refer to modulation.

\begin{table}[!ht]
\begin{center}
\begin{tabular}{|p{0.4\columnwidth}|p{0.4\columnwidth}|}
\hline
\textbf{Specification} & \textbf{Description} \\
\hline
5G Frequency Band N77  & 3800MHz-4100MHz \\ 
\hline
Carrier Bandwidth      & 100MHz \\ 
\hline
Modulation             & 256(DL)/64(UL) QAM \\ 
\hline
Transmit power         & 5W per Tx path (4Tx paths) \\ 
\hline
MIMO layers            & 4x2 closed loop MIMO \\ 
\hline
TDD  (UL:DL) ratio     & 3/7  \\
\hline
\end{tabular}
\caption{5G-SA N77 network configuration}
\label{tab:5gsystem}
\end{center}
\end{table}

\begin{figure}[thpb]
\begin{center}
\includegraphics[width=0.9\columnwidth]{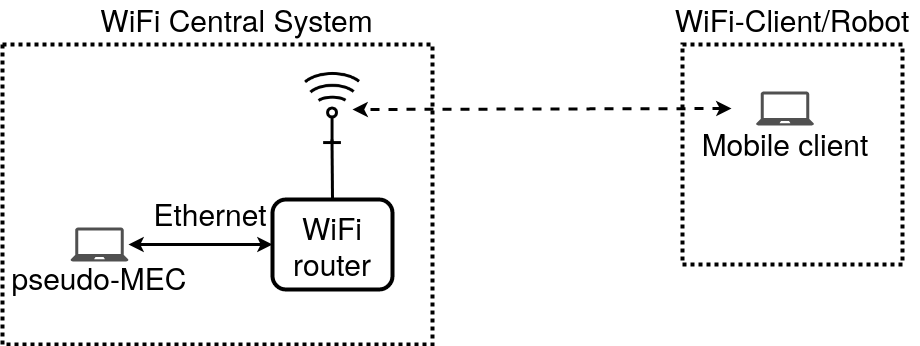}
\caption{Connection diagram for the WiFi6 network.}
\label{fig:wifi-clientdiagram}
\end{center}
\end{figure}

\item \textbf{WiFi6 Network}:
A connection diagram is illustrated in Figure~\ref{fig:wifi-clientdiagram}.
The pseudo-MEC, for the WiFi6 network experiments is connected via Ethernet cable (cat6) to a WiFi6 enabled router.
The mobile client on the remote controlled robot has an internal WiFi6 network card that allows it to communicate with the WiFi6 enabled router.

\paragraph{WiFi6 configuration:}
The WiFi6 network system and
relevant parameters are listed in Table~\ref{tab:wifisystem}.
Further details on the specific WiFi6 router used can be found on the manufacturer's web site\footnote{WiFi6 router - \url{https://static.tp-link.com/2021/202103/20210311/Archer AX6000(EU&US)2.0_Datasheet.pdf}}.
It should be noted that TDD is not a used feature in WiFi communication networks and \emph{QoS (Quality of Service)} groups are disabled (unassigned).
For the ideal case (highest throughput and lowest latency), the QoS feature is left disabled.
          
\begin{table}[!ht]
\begin{center}
\begin{tabular}{|p{0.4\columnwidth}|p{0.4\columnwidth}|}
\hline
\textbf{Specification}          & \textbf{Description} \\ 
\hline
5GHz Frequency Band (802.11ax)  & 5160-5895MHz \\ 
\hline
Carrier Bandwidth               & 40-160MHz \\
\hline
Modulation                      & (up to) 1024(DL/UL) QAM \\ 
\hline
Transmit power                  & 1W \\
\hline
TDD (UL:DL) ratio               & N/A \\
\hline
\end{tabular}
\caption{WiFi6 central router configuration}
\label{tab:wifisystem}
\end{center}
\end{table}

\begin{figure}[thpb]
\begin{center}
\includegraphics[width=0.9\columnwidth]{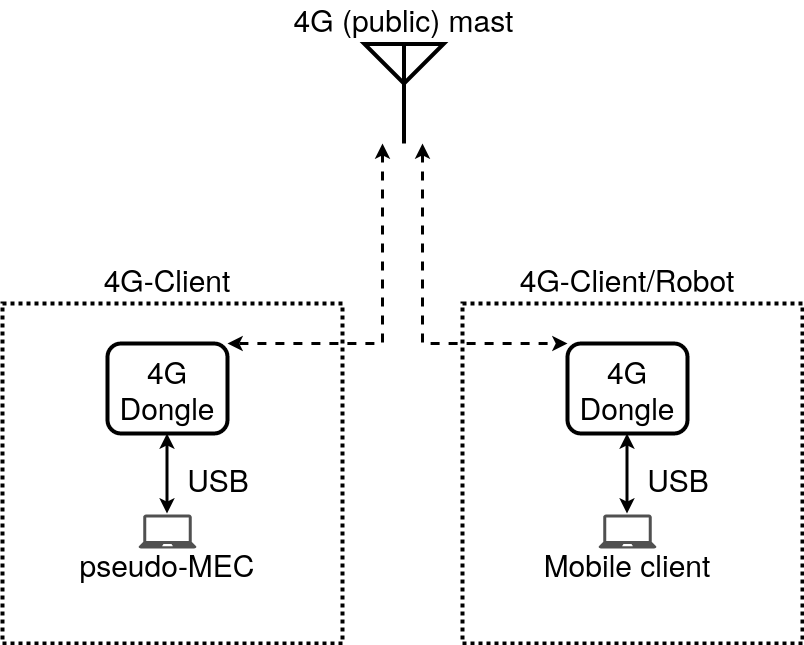}
\caption{Connection diagram for the 4G network.}
\label{fig:4g-clientdiagram}
\end{center}
\end{figure}

\item \textbf{4G Network:}
Because we used a public 4G network, the pseudo-MEC could not be directly connected to a receiving 4G mast.
Instead, the pseudo-MEC and mobile client analogy for the 4G experiments is replaced by a client-to-client analogy, illustrated in Figure~\ref{fig:4g-clientdiagram}.
Both mobile clients used for 4G network experiments, connect to the network via external USB dongle devices.
The external device used for 4G networking is the D-Link DWM-222\footnote{D-Link (4G dongle) - \url{https://www.dlink.com/en/products/dwm-222-4g-lte-usb-adapter}}.
            
\paragraph{4G configuration:}
The D-Link DWM-222 supports any UK commercial SIM card network carrier and connects to any device via USB2.0 connection.
The maximum data rate of USB2.0 is 480Mbps, which is enough to test the maximum theoretical speed of 4G communication, which is 300Mbps DL and 150Mbps UL.
However, actual 4G commercial download and upload speed is approximately 6\% (18.4Mbps) and 10\% (14.7Mbps) of the theoretical maximums (achieved by EE\footnote{\url{https://ee.co.uk/}}), respectively.
The UL/DL data is aggregated from over 210,000 mobile phones across the UK~\citep{Ofcom}.
To demonstrate the maximum real-world 4G speed achieved in VLoS and within approximately 10m of a 4G mast, measurements were taken in London (UK) using Ookla\footnote{Ookla internet speed test - \url{https://www.speedtest.net/}}, which is a speed test application that downloads and uploads a short burst of data to measure throughput;
results were 100Mbps DL and 20Mbps UL.
In a normal usage scenario, it is extremely unlikely to achieve such 4G speeds, as this would require a user to be in close proximity, in VLoS and be able to predict low network traffic load for a particular public 4G mast, and additionally know if the server of the service they want to use is spatially close (fewer hops between network nodes to reach the server) and that it employs state-of-the-art network capabilities.
Table~\ref{tab:4gsystem} lists all the known parameters for the 4G network. 
It should be noted that the actual TDD ratio is unknown and usually dynamic depending on the 4G network carrier.
However, the maximum theoretical speeds and real-world practical speeds are well known and documented for 4G, these are given in Table \ref{tab:4gsystem}.
          
\begin{table}[!ht]
\begin{center}
\begin{tabular}{|p{0.4\columnwidth}|p{0.4\columnwidth}|}
\hline
\textbf{Specification}     & \textbf{Description} \\
\hline
LTE Frequency Band         & 800MHz-2600MHz \\
\hline
Carrier Bandwidth          & 1-20MHz \\
\hline
Modulation                 & 256(DL)/64(UL)QAM \\ 
\hline
Transmit power             & 0.2W \\
\hline
UL:DL (in Mbps) & 150:300(theoretical) 20:100(real-world) \\
\hline
\end{tabular}
\caption{4G D-Link configuration~\citep{dlink:datasheet}}
\label{tab:4gsystem}
\end{center}
\end{table}
          
\end{itemize}

\subsection{Tunnelling 5G Communication}
\label{sec:method:tunnel}

\begin{figure}[thpb]
\begin{center}
\includegraphics[width=0.9\columnwidth]{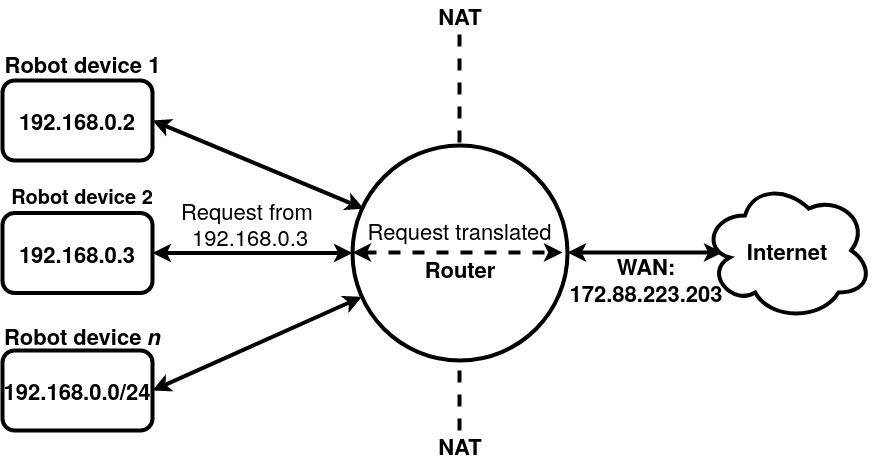}
\caption{A simplified diagram showing how messages are handled by a router using NAT.}
\label{fig:nat-diagram}
\end{center}
\end{figure}

Tunnelling is a network protocol that allows the secure transmission of private data over a public network.
It is a way of giving users of a public network access to network resources that they would not otherwise be able to reach\footnote{https://www.cisco.com/c/en/us/products/ios-nx-os-software/tunneling/index.html}.
In some rare instances, tunnelling is used to enable unsupported network protocols and to bypass firewalls.
The nature of the private 5G network and public 4G network experiments required us to use tunnelling for this purpose.
The current 5G network setup uses a network address translation (NAT) layer, which hides any connected devices' IP for better security.
For research use cases and experimentation, the NAT layer presents an issue as it makes direct communication between connected devices impossible.
The way of circumnavigating the issue is by creating a private tunnel connection between directly communicating devices, which is what has been done for the experiments described in Section~\ref{sec:physical:exp}.
In the future, NAT forwarding will be enabled as a feature for the private 5G network to allow direct communication without the need for tunnelling.
However, for public 4G network experiments, removing the NAT layer is not an option as it is controlled by the network carrier and security is a very important and concerning issue on public networks.
Thus, it will always be necessary to bypass the security measures put on public networks and to enable certain network protocols to run between the pseudo-MEC (server) and remote-controlled robot (client).
The public 4G network results presented in Sections~\ref{sec:physical:exp} and~\ref{sec:sim:exp} are used to demonstrate the best possible communication with current commercial technology in rural areas, which many farmers currently contend with\footnote{Depending on the location of farm fields and what type of mobile network is available.}.
4G is used as the ``benchmark to beat'' for 5G, while WiFi6, although restrictive in its use case in agriculture, is used to show how close 5G gets to a state-of-the-art wireless local area network (WLAN). 
A simplified network diagram in Figure~\ref{fig:nat-diagram} shows how NAT works for the 4G and 5G networks.

Unlike mobile networks, e.g. 3G, 4G, 5G, etc., WiFi, and by extension WiFi6, routers do not need to hide wireless local devices' IP addresses.
The functionality of NAT is usually required only when a device is connected to the internet (online), which is not always required by farmers.
If a WiFi enabled network is required to upload data to the cloud or to an online server, this can be done without introducing NAT to the \textit{local} wireless area network. 
However, introducing an online component to WLANs can cause bottlenecks to occur due to low throughput capabilities of a specific internet service provider (ISP) or geographical area.

To achieve bidirectional communication for 4G and 5G, a peer-to-peer \emph{tunnelling} network service was created using WireGuard~\citep{donenfeld:wire}.
Network tunnelling can increase delay if the network path taken between communicating devices is not direct, i.e. requests have to be made to the virtual private network (VPN) or tunnelling service; in addition, communication paths can take unknown hops to reach a destination. 
However, the tunnelling network for the private 5G network is made up of only 2 end-point laptops, which means that there is minimal delay in the system.
For example, Donenfeld~\citep{donenfeld:wire} performed tests using ideal conditions (2 end-point devices connected with an Ethernet cable), and WireGuard achieved the lowest ping time against all other tested applications, with a latency of \textasciitilde0.403ms.
A WireGuard experiment over-the-air cannot be conducted accurately enough as dynamic environment conditions and distance to the 5G mast (14.5 m above ground) are hard to measure precisely and are highly variable.
However, from the latency results shown in Section~\ref{sec:exp1:results}, any delay introduced by WireGuard is considered insignificant.

To allow for two devices to directly communicate over a public network, a different type of WireGuard service is required, i.e. server-client tunnelling network.
For example, the public 4G network experiments were configured using a WireGuard server-client tunnelling network to bypass the ISP gateway (anonymity) that comes with standard public wireless communications.
However, this means that there is an increase in WireGuard delay path routing and it is more complex to calculate true \emph{Round-Trip Time (RTT)} latency.

\section{Physical Experiments: Network Throughput and Latency}
\label{sec:physical:exp}

\begin{figure}[thpb]
\begin{center}
\includegraphics[width=0.7\columnwidth]{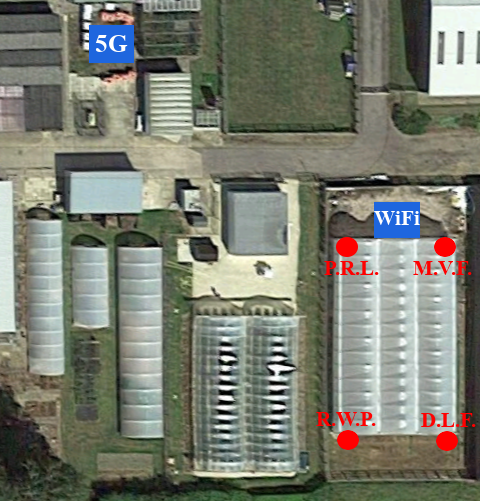}
\caption{Satellite image showing the four experiment locations, identified using abbreviated what3words, in the Vegetable Polytunnel.}
\label{fig:newpoly}
\end{center}
\end{figure}

\begin{figure}[thpb]
\begin{center}
\includegraphics[width=1.0\columnwidth]{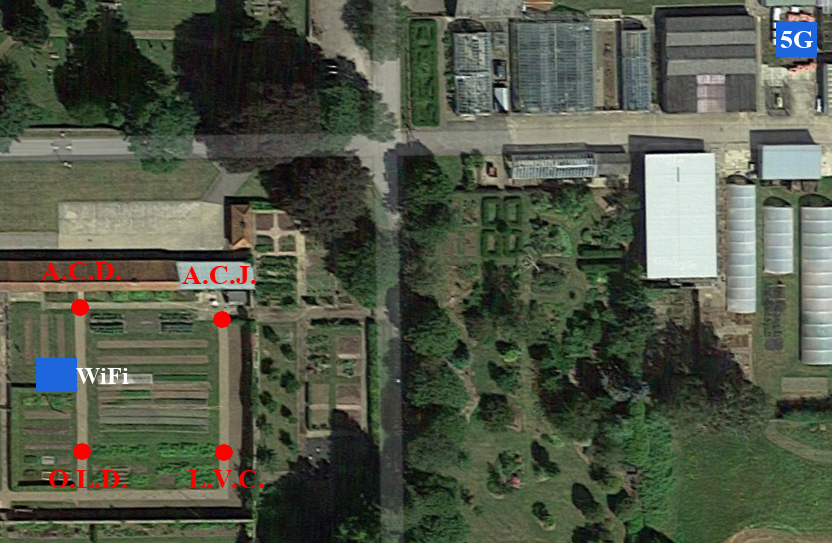}
\caption{Satellite image showing the four experiment locations, identified using abbreviated what3words, in the Walled Garden.}
\label{fig:walledgarden}
\end{center}
\end{figure}

Wireless network experiments were conducted in four corners of two test environments, aforementioned in Section~\ref{sec:method:locations}, the Vegetable Polytunnel and the Walled Garden.
In total, experiments were conducted in 8 geographically different points and, at each point, an experiment lasted 30 seconds and was repeated 5 times.
This allows for results to be interpreted in two ways:
firstly, by taking the results of each experiment run of 30 seconds separately;
and
secondly, by combining the results of 5 experiments over 30 seconds, which totals 2 minutes and 30 seconds of acquired data per location.
The results presented in Section~\ref{sec:exp1:results} are obtained using the first methodology.
The commercial mapping tool \emph{What3Words (W3W)}\footnote{\url{https://what3words.com}} was used to identify and mark the 8 data collection points where experiments were conducted, illustrated in Figures~\ref{fig:newpoly} and~\ref{fig:walledgarden}.
For ease of visualising the results in Section \ref{sec:exp1:results}, the points used for data collection are labelled with the first letter of each W3W specification, and the core network (access point) for 5G and WiFi6 are labelled. 
The approximate distances between each data collection point and the access points (5G and WiFi6) are given in Tables \ref{5g-wifi-poly-tab} and \ref{5g-wifi-wg-tab}.
Finally, the physical experiments and results are briefly discussed in \ref{sec:exp1:discussion}.

\begin{table}[ht]
\begin{center}
\begin{tabular}{|ccc|}
\hline
\multicolumn{3}{|c|}{\textbf{Vegetable Polytunnel}} \\ \hline
\multicolumn{1}{|c|}{\textbf{W3W Location}} & \multicolumn{2}{c|}{\textbf{Distance(m)}} \\ \hline
\multicolumn{1}{|c|}{\textbf{}} & \multicolumn{1}{c|}{\textbf{5G}} & \textbf{WiFi} \\ \hline
\multicolumn{1}{|c|}{P.R.L.} & \multicolumn{1}{c|}{49.1} & 8.3 \\ \hline
\multicolumn{1}{|c|}{M.V.F.} & \multicolumn{1}{c|}{61.5} & 8.6 \\ \hline
\multicolumn{1}{|c|}{R.W.P.} & \multicolumn{1}{c|}{72.0} & 32.6 \\ \hline
\multicolumn{1}{|c|}{D.L.F.} & \multicolumn{1}{c|}{81.4} & 32.9 \\ \hline
\end{tabular}
\caption{The distance between 5G/WiFi access point and the data collection points in the Vegetable Polytunnel.}
\label{5g-wifi-poly-tab}
\end{center}
\end{table}
\begin{table}[ht]
\begin{center}
\begin{tabular}{|ccc|}
\hline
\multicolumn{3}{|c|}{\textbf{Walled Garden}} \\ \hline
\multicolumn{1}{|c|}{\textbf{W3W Location}} & \multicolumn{2}{c|}{\textbf{Distance(m)}} \\ \hline
\multicolumn{1}{|c|}{\textbf{}} & \multicolumn{1}{c|}{\textbf{5G}} & \textbf{WiFi} \\ \hline
\multicolumn{1}{|c|}{A.C.D.} & \multicolumn{1}{c|}{143.4} & 14.0 \\ \hline
\multicolumn{1}{|c|}{A.C.J.} & \multicolumn{1}{c|}{119.5} & 32.2 \\ \hline
\multicolumn{1}{|c|}{O.L.D.} & \multicolumn{1}{c|}{154.8} & 14.4 \\ \hline
\multicolumn{1}{|c|}{L.V.C.} & \multicolumn{1}{c|}{132.3} & 33.2 \\ \hline
\end{tabular}
\caption{The distance between 5G/WiFi access point and the data collection points in the Walled Garden.}
\label{5g-wifi-wg-tab}
\end{center}
\end{table}

\subsection{Performance Metrics}
\label{sec:exp1:metrics}

To test network stability and performance, different video streaming settings were used, namely 1-RGB, 4-RGB and 1-RGBD video streams\footnote{All video stream data in experiments was compressed.}.
The number at the start of RGB denotes the number of video streams, for example 1-RGB denotes \textit{one} RGB video stream.
The in-field robot streams video data back to the pseudo-MEC.
The 1-RGB video stream experiment tests realistic latency conditions in what can be considered \textit{typical} or \textit{medium} network load.
The 4-RGB video stream experiment is used to test how 4G, 5G and WiFi6 deal with multiple data streams communicating at the same time.
Moreover, four video streams can be considered \textit{heavy} network load, which is expected to increase latency for all network types.
Finally, the 1-RGBD stream experiment is used as a method to analyse how the different networks react to a single source of consistent \textit{heavy} network load.
However, it is important to stress that it was never the intention of the authors to analyse maximum throughput or lowest latency, but rather to demonstrate the practical results and to evaluate the performance of state-of-the-art network systems, i.e., 5G and WiFi6, and a commonly used commercial network system, i.e., 4G.

There were three independent variables in the network experiments: 
\textit{location}, 
\textit{network type} 
and 
\textit{video stream number}.
There were two dependent variables, i.e. the raw data collected to assess performance:
\emph{latency}, measured in \emph{microseconds~(ms)}, and
\emph{throughput}, measured in \emph{Megabits per second~(Mbps)}.
The results are presented next.

\subsection{Results}
\label{sec:exp1:results}

While a larger set of performance metrics were collected during the experiments described in this paper,
a selected portion of the results that best 
illustrate our aims
are reported here.
For the two performance metrics, three statistics are presented: 
mean, standard deviation and minimum
\textit{latency~(ms)}
and
mean, standard deviation and maximum
\textit{throughput~(Mbps)}.

Throughput\footnote{Wherever ``\textit{throughput}'' results are shown or discussed they depict ``\textit{data sent}'' performance metric.} results are interpreted from the point of view of the in-field mobile robot, i.e., data sent.
The \textit{data sent} metric is much higher in proportion to the \textit{data received} from the pseudo-MEC, which is negligible.
This is because the mobile robot receives basic network telemetry data, video stream control messages to start and stop a stream, and co-ordinate information identifying weed locations in an image, as a consequence it is not investigated in this work.

\begin{figure}[ht]
\begin{center}
\begin{tabular}{c}
\includegraphics[width=0.9\columnwidth]{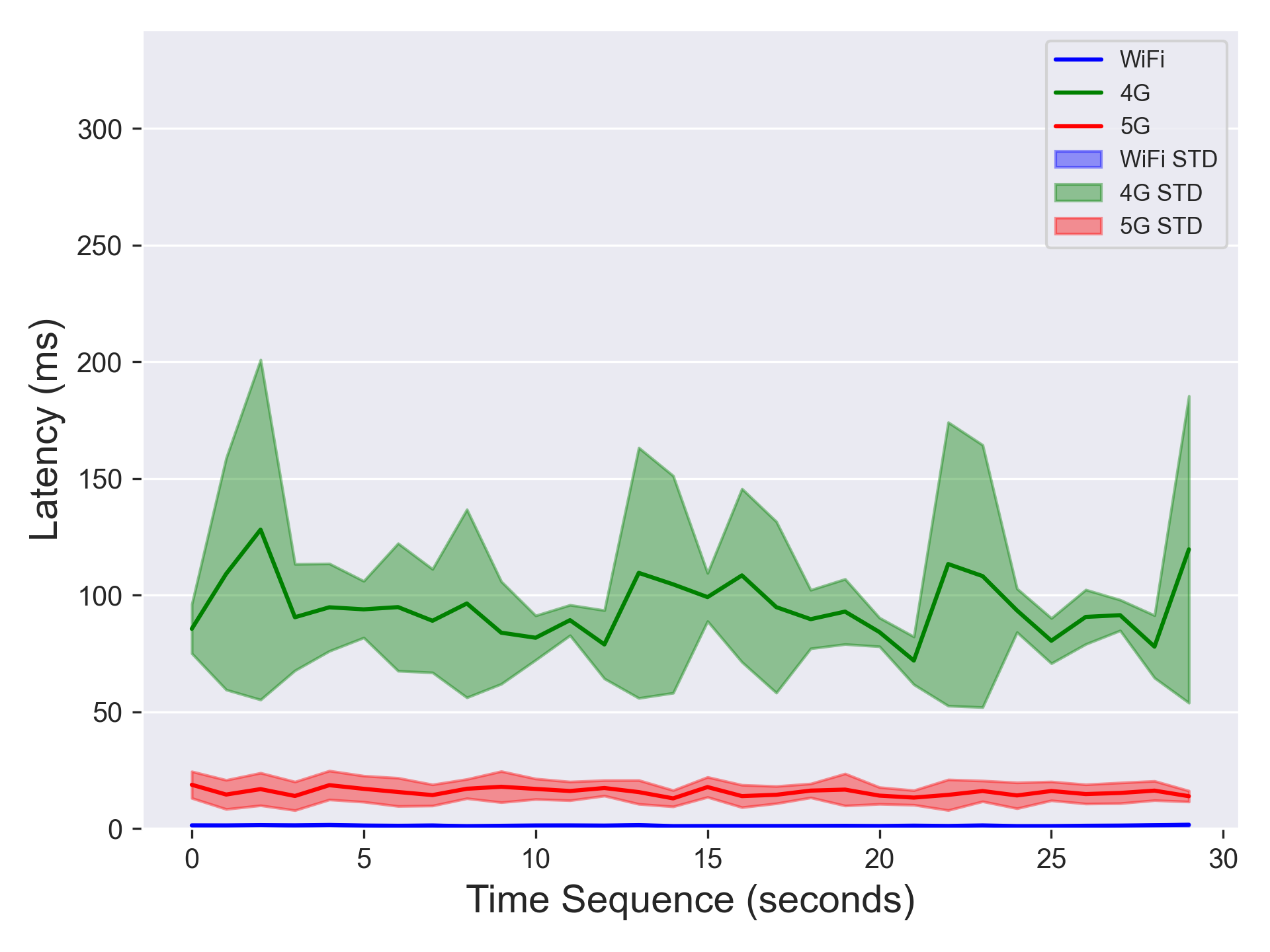} \\
(a) Vegetable Polytunnel\\
\includegraphics[width=0.9\columnwidth]{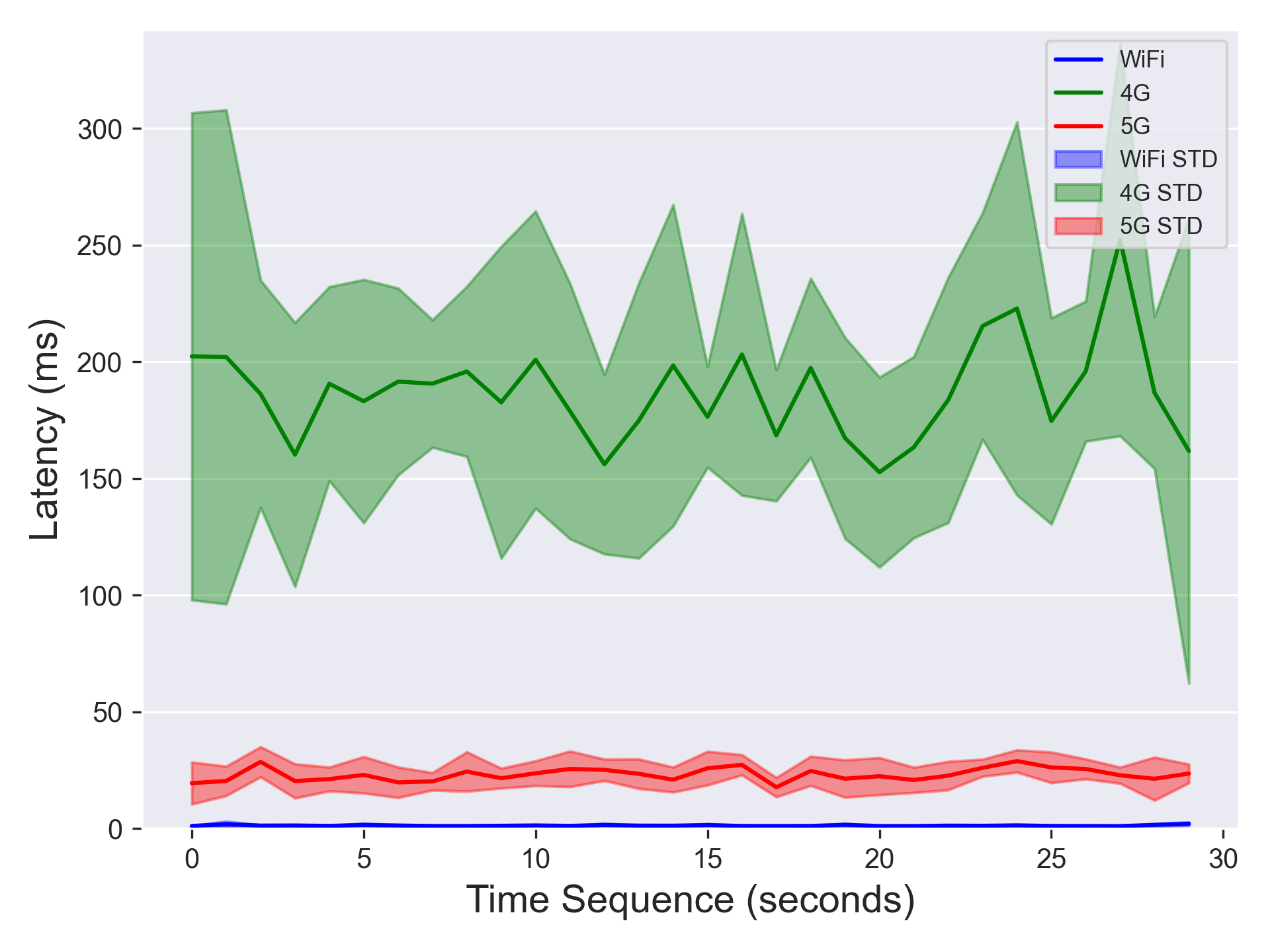} \\
(b) Walled Garden\\
\end{tabular}
\caption{Best network latency results, averaged over 5 experimental runs gathered from a single ``best'' location (please refer to Table \ref{tab:results}). The mean is the solid line in the centre of the shaded regions, which shows $\pm~1$ standard deviation.
}%
\label{fig:latency-results}%
\end{center}
\end{figure}

\begin{figure}[ht]
\begin{center}
\begin{tabular}{c}
\includegraphics[width=0.9\columnwidth]{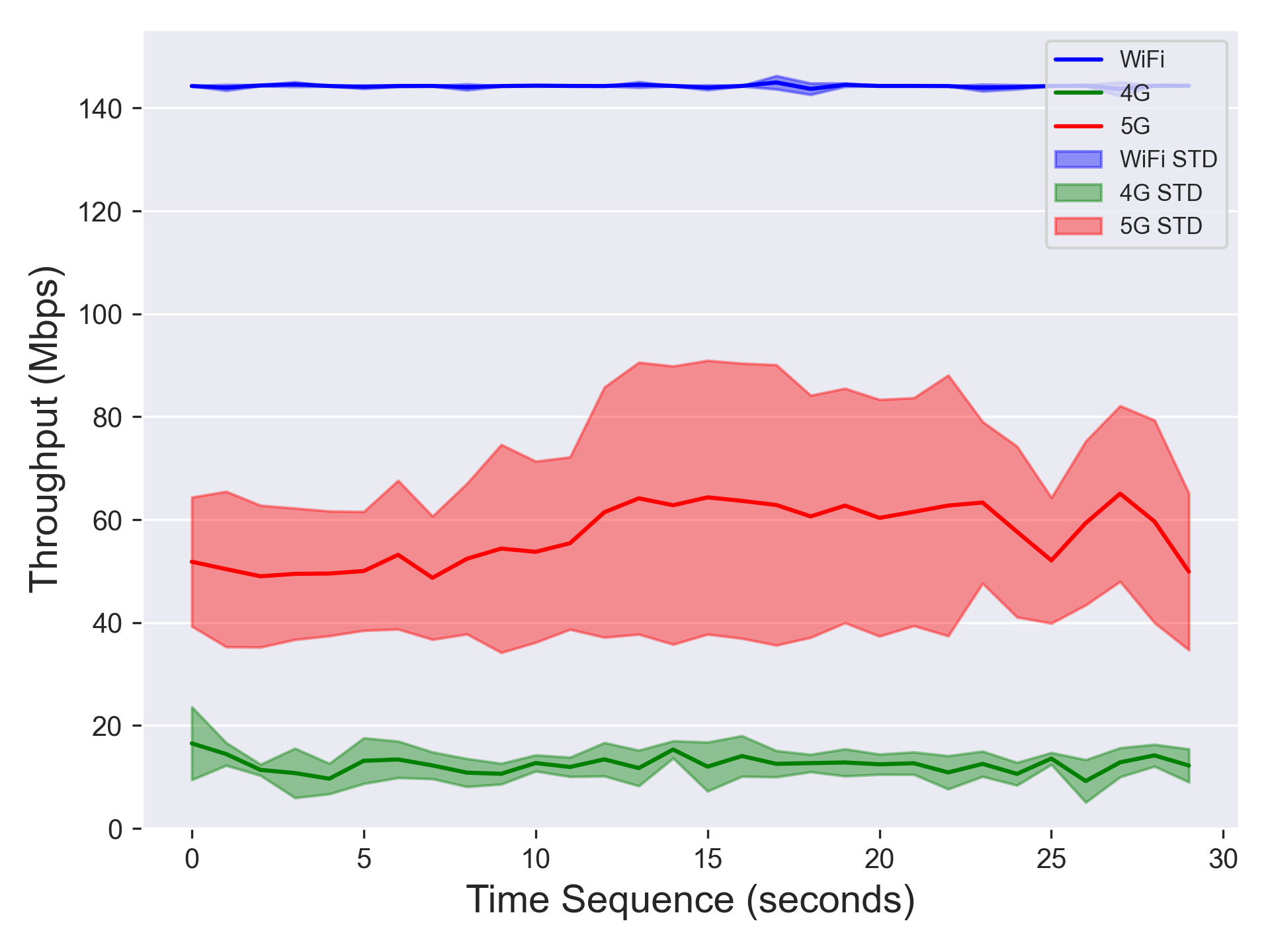} \\
(a) Vegetable Polytunnel\\
\includegraphics[width=0.9\columnwidth]{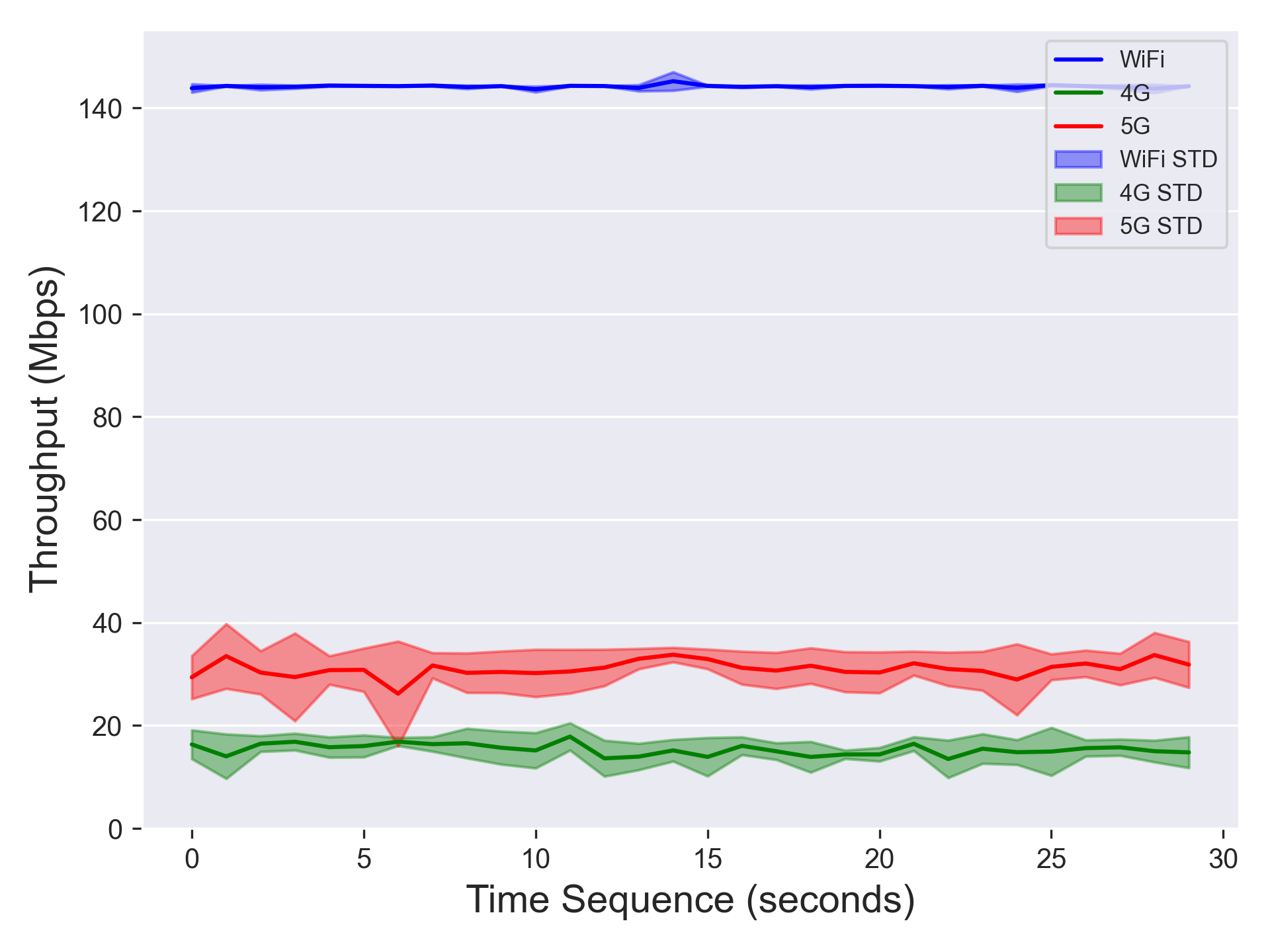}\\
(b) Walled Garden\\
\end{tabular}
\caption{Network throughput results, averaged over 5 experimental runs gathered from a single ``best'' location (please refer to Table \ref{tab:results}). The mean is the solid line in the centre of the shaded regions, which shows $\pm~1$ standard deviation.
}%
\label{fig:throughput-results}%
\end{center}
\end{figure}

The data collection point (geographical point) with the best results for latency (lowest mean and minimal latency) and throughput (highest mean and maximum throughput)
is selected for each of the two environments and shown in Table~\ref{tab:results}\footnote{For a complete view of all data collection points covering the six key performance metrics, refer to Appendix \ref{appendix}}.
Figure \ref{fig:latency-results} and Figure \ref{fig:throughput-results} visually show the data from Table~\ref{tab:results} for each of the environments, Walled Garden and Vegetable Polytunnel.
The wireless networks' latency results ordering, in Figure \ref{fig:latency-results}, remained the same throughout all data collection points in both test environments.
It was always the case that WiFi6 had the lowest latency followed by 5G, whereas 4G had the highest latency which was ten times higher than the latter.
The ordering of the wireless networks' performance remained similar for throughput, as shown in Figure \ref{fig:throughput-results}.
In all instances WiFi6 outperformed 5G and greatly outperformed 4G. 
Whereas, 5G outperformed 4G in all environments.
Finally, the distance between the two environments from each access point is averaged and compared for 5G and WiFi6.
The 5G mast is an average distance of 66.0 metres and 137.5 metres from the Vegetable Polytunnel and Walled Garden, respectively. Whereas, the WiFi6 router is an average distance of 20.6 metres and 23.5 metres from the Vegetable Polytunnel and Walled Garden, respectively.
A difference of 45.4 metres and 114.0 metres respectively, between the two corresponding environments and wireless networks.

\begin{table}[ht]
\begin{center}
\begin{tabular}{|r|r|r|r|r|r|r|}
\hline
\multicolumn{7}{|c|}{\textbf{Vegetable Polytunnel}} \\ \hline
\textbf{Network} & 
\multicolumn{3}{c|}{\textbf{Latency (ms)}} & 
\multicolumn{3}{c|}{\textbf{Throughput (Mbps)}} \\
\textbf{Type} & 
\textbf{Mean} & 
\textbf{Min} &
\multicolumn{1}{c|}{\begin{tabular}[c]{@{}c@{}}\textbf{Loc-}\\ \textbf{ation}\end{tabular}} &
\textbf{Mean} &
\textbf{Max} &
\multicolumn{1}{c|}{\begin{tabular}[c]{@{}c@{}}\textbf{Loc-}\\ \textbf{ation}\end{tabular}} \\
\hline
4G    & 94.9 & 72.0 & DLF & 12.5 &  16.5 & DLF \\ \hline
WiFi6 &  1.2 &  1.0 & RWP & 144.2 & 145.0 & PRL \\ \hline
5G    & 15.7 & 10.9 & DLF & 57.1 &  65.1 & PRL \\ \hline
\hline
\multicolumn{7}{|c|}{\textbf{Walled Garden}} \\ \hline
\textbf{Network} & 
\multicolumn{3}{c|}{\textbf{Latency (ms)}} & 
\multicolumn{3}{c|}{\textbf{Throughput (Mbps)}} \\
\textbf{Type} & 
\textbf{Mean} & 
\textbf{Min} &
\multicolumn{1}{c|}{\begin{tabular}[c]{@{}c@{}}\textbf{Loc-}\\ \textbf{ation}\end{tabular}} &
\textbf{Mean} &
\textbf{Max} &
\multicolumn{1}{c|}{\begin{tabular}[c]{@{}c@{}}\textbf{Loc-}\\ \textbf{ation}\end{tabular}} \\
\hline
4G    & 187.2 & 137.6 & LVC & 15.4 &  17.8 & ACJ \\ \hline
WiFi6 &   1.3 &   1.0 & OLD & 144.2 & 149.5 & ACD \\ \hline
5G    &  23.1 &   1.0 & OLD & 31.0 & 33.8 & OLD \\ \hline 
\end{tabular}
\caption{The best results achieved in each environment for the different network types.}
\label{tab:results}
\end{center}
\end{table}

\subsection{Discussion}
\label{sec:exp1:discussion}

We show that commercially available public 4G is unrealistic to be used for high data rate and low-latency operations in the rural environment, \textit{rarely} achieving below 100ms latency and never managing to reach over 20Mbps data throughput. 
Though, it is promising that on average the public 4G network data rate is close to the actual commercial upload speeds of 14.7Mbps quoted by \cite{Ofcom}.
It should be noted that the surrounding area was thoroughly evaluated for the best 4G signal and network provider to achieve these results.
However, low throughput and high latency can result in poor performance of the in-field robot, misinterpreting and mislabelling plants or even robots causing damage to plants.
A strength of public 4G networks for agriculture is that \textit{if} a rural area has any network coverage, it is quick and easy to setup with little configuration required.
However, this strength carries a weakness.
Total control and availability of the network is at the discretion of the network carrier (ISP).
Moreover, a network wide outage for the ISP means instant outage and complete disruption to normal operation on the farm.

Public 5G, which is not evaluated in this work, is \textit{expected} to perform with lower latency and higher data rate than public 4G.
Hence, it can be \textit{assumed} that public 5G can support high data rate, low-latency agri-robotics and the future smart farm.
However, this is not the case currently and it will remain so until public 5G fully matures, and even if it does, there is a chance that it will remain unrealistic like public 4G.
It needs to be considered that commercial networks do not apply a balanced TDD, i.e., more emphasis on download speed and delivering services.
Unlike a private network that can be configured to provide more balanced upload and download speeds and improve network coverage to more rural areas.
If we take a look at real-world data provided by Ookla~\citep{ookla:5g} for Q1-Q2 of 2021, the highest 5G \textit{upload} data achieved is 41.79Mbps by South Korea.
South Korea have been the leader in network technology and internet infrastructure since the late 90s early 00s~\citep{heejin:southkorea-networking} and they are world leading in 5G as well~\citep{massaro:southkorea-5G}.
Yet, the remaining bottleneck for public 5G seems to be upload speed.
Getting over the maturity and configuration hurdle, the lack of control over the network and relying on an ISP, as is the case for 4G, remains an issue.

The private 5G available at the University of Lincoln has proved why it is better than public 5G, by showing greater upload speeds achieved in real-world experiments of 57.1Mbps with VLoS and 31.0Mbps with NVLoS, Table~\ref{tab:results}.
The slowest average upload speed is approximately double that of the UK average according to \cite{ookla:5g}.
Upload speeds over 30Mbps can support at least one live video stream and bi-directional communication and 60Mbps can support two live streams and bi-directional communication.
Moreover, the latter case can support multiple live streams, however video streaming will not be real-time and will not be running at 30FPS.
The private 5G 4-RGB streaming experiments showed significant reduction in video stream quality and speed, with some streams buffering for a few seconds before starting back up again.
The fact that four video streams shared bandwidth meant that the system was trying to balance resources and all four streams were not running at the same speed, i.e. some smoother than others.
Whereas, the 1-RGBD stream experiment experienced \textit{slowness} or \textit{choppiness} and was not running at 30FPS.
The expected bandwidth requirement for live RGBD video streaming is \textasciitilde145.0Mbps, 5G could support approximately half the required bandwidth.

The private WiFi6 (local) network was evaluated as it has recently become commercially available and it is state-of-the-art in terms of network features and performance, introducing higher network speeds and very low-latency.
It was expected that WiFi6 will beat 5G in data throughput, and in fact it leads 5G by \textasciitilde2.5 times in upload data speeds.
WiFi6 unexpectedly beats 5G in latency time as well, by being as much as \textasciitilde13 times lower.
However, the distances at which these results are obtained are not the same as for 5G, and the NVLoS experienced by 5G is not present for WiFi6.

The point with the greatest distance for WiFi6 is 33.2 metres and for 5G is 154.8 metres, over 4.5 times greater for the latter.
The attenuation of a WiFi signal is exponential and at a distance greater than 100 metres there would be no signal (communication) at all.
High gain antenna could be used to boost WiFi signal, however such antennae do not exist commercially for WiFi6.
Moreover, a license needs to be obtained to operate such antennae for the WiFi standard making it very likely the case that WiFi6 will also require license to operate signal boosting antennae.

WiFi6 results are demonstrably better than 5G and at a completely different level compared to 4G, however there are many situations where WiFi6 is not the best option in agriculture.
For example, the experiments conducted in this work used \textit{only} the WiFi6 standard, and support was disabled for older WiFi standards.
This forced all devices to use the latest standard for message transmission ensuring lowest possible latency and highest throughput.
However, in practical environments (i.e., farm) it can be beneficial to enable multi-WiFi support, allowing certain sensors to use older standards, which may allow for greater compatibility, coverage and more robust signal-strength to distance drop off (better attenuation at greater distances).
Moreover, not many discrete and low power WiFi6 network devices exist on the market.
Most sensors used by agronomists or farmers for monitoring rainfall, soil moisture, light levels, etc., do not support WiFi6.
Because of this, WiFi6 is less known and not many real world use cases and data exist yet.

\section{Simulation Experiments: Real-Time Operation and Control} 
\label{sec:sim:exp}

A future technology being introduced to 5G is \emph{ultra-reliable low-latency communication (URLLC)} that will guarantee $\sim$99.999\% reliability of communication and real-time low-latency.
This feature should have been available for FR1 (Frequency range 1 - i.e., 5G N77 band) in early 2021, but its release has been delayed by most system providers, including the private 5G network at the University of Lincoln.
URLLC~\citep{sachs:5gurllc} is a 5G feature that has been marked to bring realisation to many technologies, one of which is \emph{V2X} (Vehicle-to-Everything) networks, designed to provide real-time reliable communication to assist navigation in fully autonomous vehicles, traffic control and road safety protocols~\citep{shanzhi:v2x,rashid:ti-urllc}.
Sachs et al.~(\citeyear{sachs:5gurllc}) explain that the theoretical worst-case transmission latencies differ depending on network configuration, showing that RTT latency can range from as low as \textasciitilde0.8ms to as high as \textasciitilde6.3ms, depending on configuration.
However, according to the official 3GPP technical specification~\citep{3gpp.38.913}, the intended theoretical RTT latency target for URLLC is 1ms. 

To not convolute the simulated experiment results, a comparison and evaluation is given assuming the theoretical URLLC value (1ms) given in the \cite{3gpp.38.913} report.
Moreover, we do not evaluate the reliability of the wireless networks, as none of them, including the current private 5G network, have the URLLC feature available.
URLLC is not a feature that exists for 4G or WiFi6, and as mentioned it is not available for most 5G systems.
The results in this work are \textbf{not} intended to directly challenge or prove 5G URLLC, furthermore we do acknowledge that there are targeted use cases for this feature that 4G and WiFi6 cannot support, e.g. V2X, due to network and infrastructure limitations.
This simulation strictly compares the RTT latency of the three wireless networks against the URLLC theoretical specification to primarily investigate the real-time speed-up of robot operation in the field and the improved performance.

\subsection{Experiment Design}
\label{sec:exp2:design}

The objective of this experiment is to analyse the ``real-time'' delay in positional accuracy between the different network types.
To perform these simulated experiments we used real world RTT mean latency results from two arbitrarily chosen data collection points, \textit{P.R.L.} and \textit{R.W.P.} in the Vegetable Polytunnel, as presented in Table~\ref{tab:mean-latencies-sim}.
The 5G and WiFi6 networks in the Vegetable Polytunnel have mostly VLoS with light obstructions, e.g., metal scaffolding.
Whereas, the 4G network has some VLoS with moderate obstructions, e.g., tree lines, metal scaffolding and general RF interference that can occur over longer distance communication.

The approximate distance between point \textit{P.R.L.} and \textit{R.W.P.} is \textasciitilde30 m.
Overall, two separate simulation experiments are conducted.
In each experiment the RTT mean latency result is taken from one of the points (i.e. \textit{P.R.L.}/\textit{R.W.P.}) and it is used to simulate the accumulated delay experienced by the remote controlled robot for each metre of travel.


\begin{table}[!ht]
\begin{tabular}{|c|ccc|}
\hline
Location & \multicolumn{3}{c|}{Mean Latency (ms)} \\ \hline
 & \multicolumn{1}{c|}{5G} & \multicolumn{1}{c|}{WiFi6} & 4G \\ \hline
P.R.L. & \multicolumn{1}{c|}{29.5} & \multicolumn{1}{c|}{1.2} & \multicolumn{1}{c|}{216.7} \\ \hline
R.W.P & \multicolumn{1}{c|}{22.9} & \multicolumn{1}{c|}{1.3} & \multicolumn{1}{c|}{294.0} \\ \hline
\end{tabular}
\caption{The mean latency results for points P.R.L. and R.W.P. for each of the three wireless networks.}
\label{tab:mean-latencies-sim}
\end{table}

\begin{figure}[ht]
\begin{center}
\begin{tabular}{c}
\includegraphics[width=0.9\columnwidth]{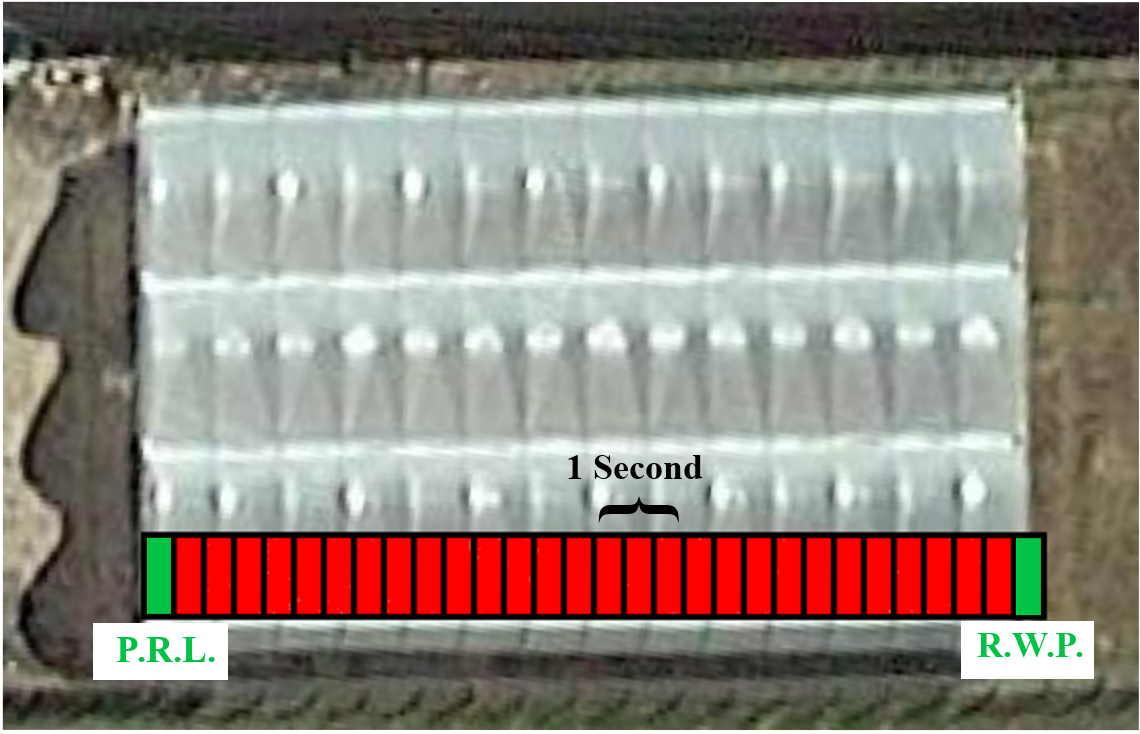} \\
(a) Satellite image of the Vegetable Polytunnel (rotated 90$\degree$ \\
anti-clockwise).\\
\includegraphics[width=0.9\columnwidth]{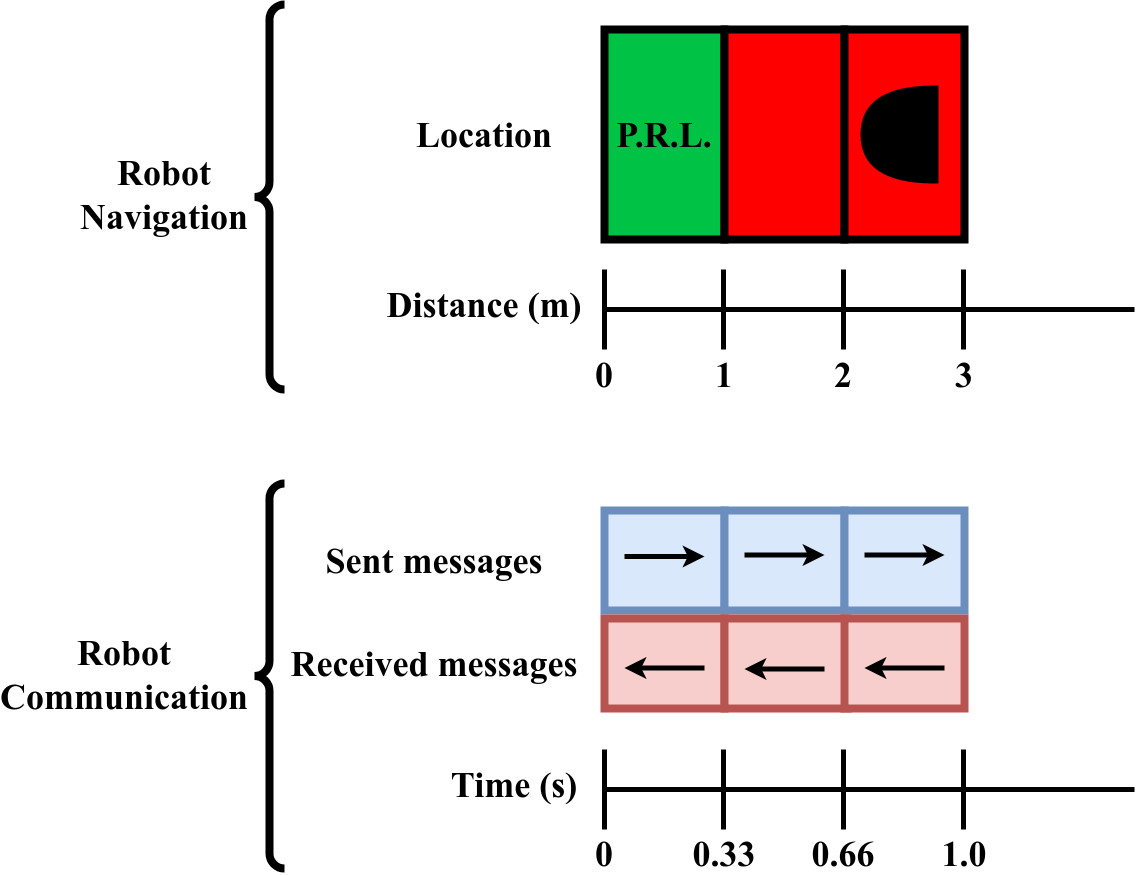}\\
(b) Magnified overview of robot navigation and (instant) \\
communication over a period of 1 second. \\
\end{tabular}
\caption{Image (a) shows 30 location spaces each depicting 1 metre, between data points \textit{P.R.L.} and \textit{R.W.P.}, representing the simulated path of the robot. Image (b) shows the magnified operation of the simulated robot if it had instantaneous (ideal) communication.}%
\label{fig:localisation-exp}%
\end{center}
\end{figure}

To demonstrate real-time positional accuracy, for every metre that the simulated remote-controlled robot moves, its location is updated and sent to the pseudo-MEC (remote server) and a processed reply message is sent back.
It is approximated that points \textit{P.R.L.} and \textit{R.W.P.} are 30 metres apart, therefore 30 location steps are generated as shown in Figure \ref{fig:localisation-exp}(a).
The robot is set to move with a velocity of 3 $m.s^{-1}$, which means that every second, 3 location spaces are passed.
At the same time, 3 location messages are sent to the pseudo-MEC and 3 command messages (e.g., spray, collision avoidance, GPS data, etc.) are received by the simulated robot every second.
We can ignore the payload (size) of location messages and command messages altogether as they are negligibly small.
Moreover, we will conceptualise that the pseudo-MEC already has the most up-to-date image data stored for each location along the path of the remote controlled robot, which means that expensive image data is not transmitted during these experiments.
Thus the most important element of the experiment is the transmission of messages.
For every metre the robot moves, one location update message is sent and one weed location message (bounding box) is received, as described in Figure \ref{fig:localisation-exp}(b) and accompanying Table \ref{tab:robot-params}.

The image processing pipeline described in Section~\ref{sec:method:ML} can process images at speeds as fast as \textasciitilde14.5ms per image.
However, in this simulated experiment, to illustrate our point more clearly, it will be assumed that the pseudo-MEC will be processing more complex images.
Therefore, the pseudo-MEC's speed of processing will be taken to be the same as the average time it takes a human to react in real-time to a sudden change on screen.
This processing (reaction) speed will make the simulation more conceptually easy to comprehend.
This value is
assumed fixed and independent of task type, 
and is set to 273ms, the median human hand-eye reaction time~\citep{latency:humanbench}.
To further simplify the simulated experiments, robot velocity is assumed fixed and other external factors contributing to latency are ignored.
To \textit{prove} that a wireless network can support real-time operation and control, the robot in the field needs to receive weed location messages while it has not yet transitioned to a new location space.
This is vital for the correct operation of a weed spraying robot, as it needs to be able to spray the weeds correctly, while maintaining its speed.
The calculation and performance metric used to determine if a robot is still within the location space is given in the next section.

\begin{table}[ht]
\centering
\begin{tabular}{|cc|}
\hline
\multicolumn{1}{|c|}{Fixed robot velocity}                  & 3 $m.s^{-1}$   \\ \hline
\multicolumn{1}{|c|}{Location update time per meter}             & 0.333 s \\ \hline
\multicolumn{1}{|c|}{Sent/Received messages per second}              & 3 $msg.s^{-1}$ \\ \hline
\multicolumn{1}{|c|}{Total messages per second}             & 6 $msg.s^{-1}$ \\ \hline
\end{tabular}
\caption{Robot Navigation and Communication Parameters}
\label{tab:robot-params}
\end{table}

\subsection{Performance Metrics and Rationale}
\label{sec:exp2:met}
As mentioned previously, sent messages will be location messages, in the form of 2D coordinate data.
Whereas, received messages can be a variety of different types of data, we will assume that it is bounding box pixel position data which identifies detected weeds in an image.
Both types of messages are sent in the form of floating point numbers, however their size is so small that it is considered negligible in terms of data throughput compared to the image data sent in the experiments in Section \ref{sec:physical:exp}.
This is desirable as we want minimal load on the network to analyse latency only.

\begin{figure}[thpb]
\begin{center}
\includegraphics[width=1.0\columnwidth]{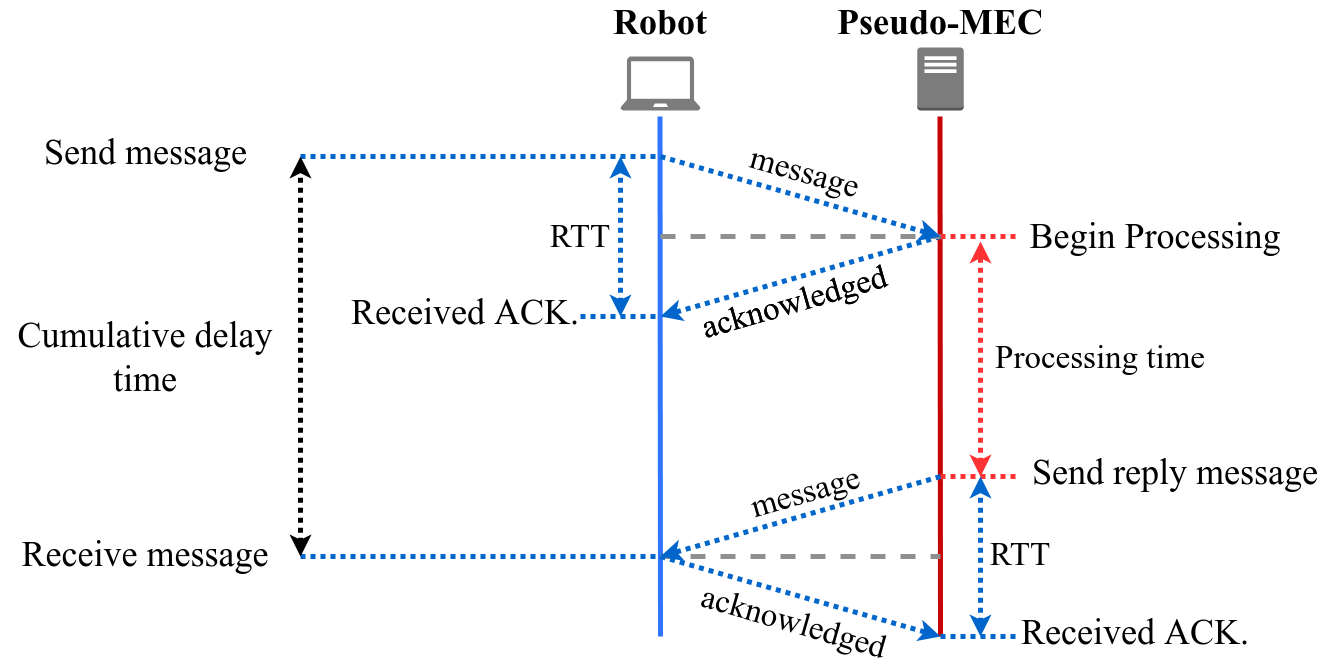}
\caption{Cumulative delay time metric.}
\label{fig:cumu-delay-time}
\end{center}
\end{figure}

From the start of the experiment, at point \textit{P.R.L.}, to the end of the experiment, at point \textit{R.W.P}, 30 location messages are sent and 30 bounding box messages are received making a total of 60 messages.
Henceforth, a sent message and the processed reply message are denoted as a \textit{pair}.
As we are making fixed assumptions for the processing time and ignoring uncertainty, the calculated \textit{cumulative delay time} for a pair is simply the RTT latency time of the network at the given location plus the processing time, as demonstrated in Figure \ref{fig:cumu-delay-time}.
The \textit{total cumulative delay time} takes into account:
(i)~the time required to send a location message from the in-field remote controlled robot to the remote server (pseudo-MEC),
(ii)~plus the time required to process the message on the server and prepare a command message in response,
(iii)~plus the time taken to send the command message from the server back to the in-field robot (denoted \textit{cumulative delay time}),
(iv)~finally, the \textit{total cumulative delay time} is the result of \textit{cumulative delay time} - \textit{location update time}, multiplied by the number of sent messages.
Therefore, \textit{total cumulative delay time} provides the delay experienced by the received messages for the total duration of travel. 

\subsection{Results}
\label{sec:exp2:results}

\begin{figure}[ht]
\begin{center}
\begin{tabular}{c}
\includegraphics[width=0.9\columnwidth]{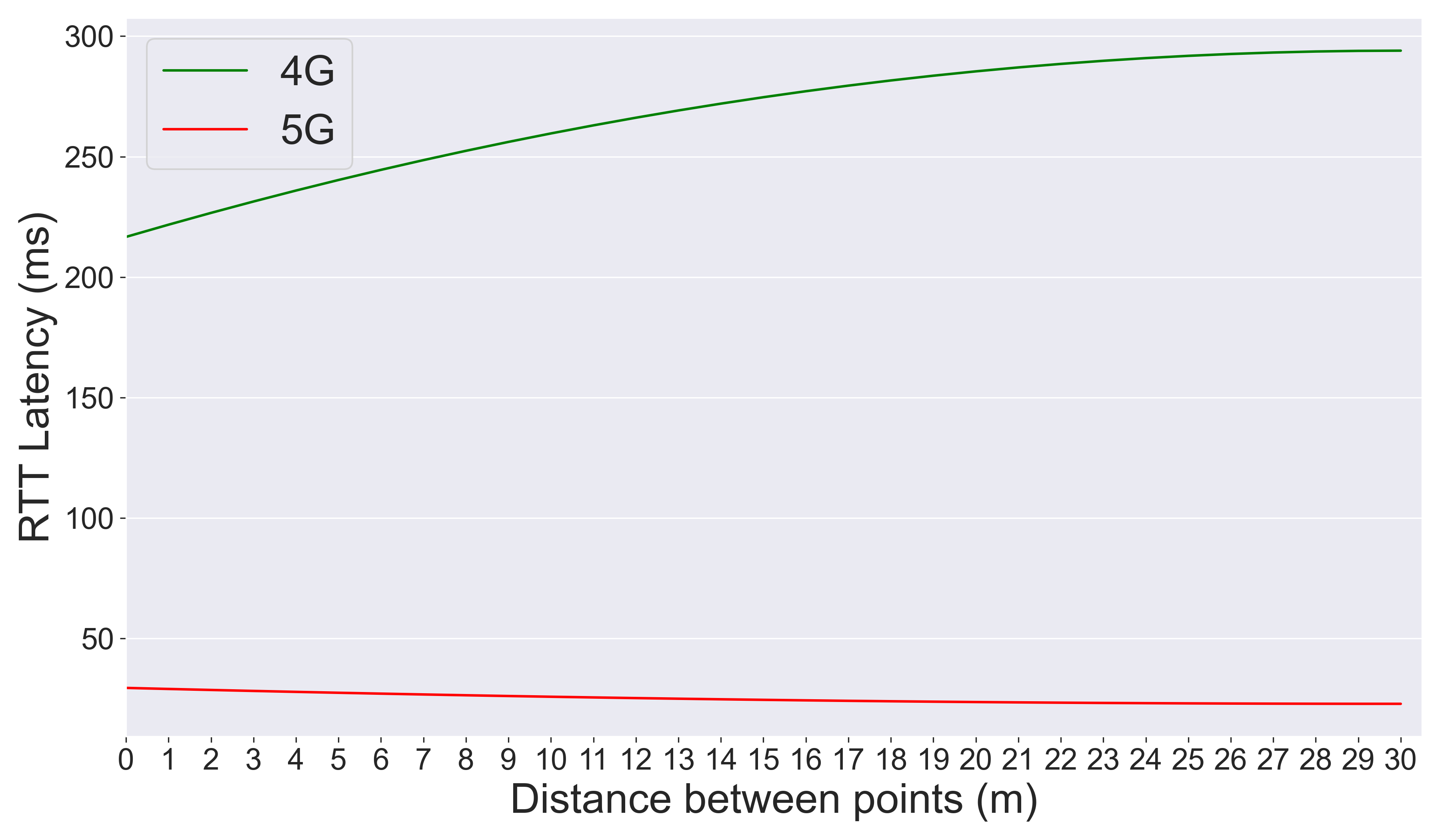} \\
(a) Trend line for 4G and 5G (lower is better). \\
\includegraphics[width=0.9\columnwidth]{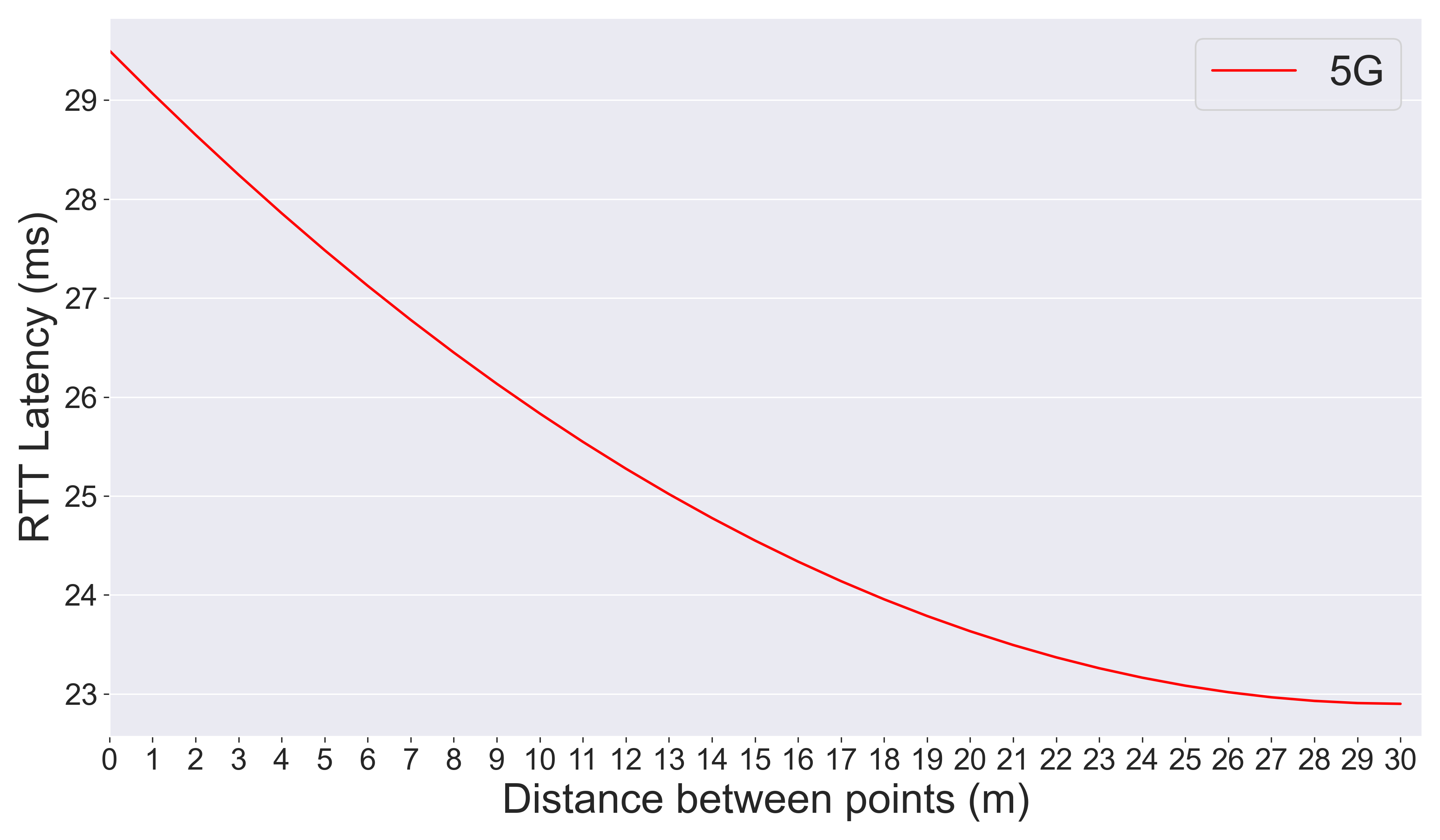}\\
(b) Trend line for 5G. \\
\end{tabular}
\caption{Image (a) shows the trend line of RTT latency as the robot moves from data point \textit{P.R.L.} to \textit{R.W.P.}. Image (b) shows a closer inspection of only the trend line of RTT latency for the 5G network.}
\label{fig:network-trendline}%
\end{center}
\end{figure}

We have RTT latency for points \textit{P.R.L.} and \textit{R.W.P.} and no real world data for the points in-between, as such we cannot perform accurate evaluation of the \textit{cumulative delay time} during the simulated navigation of our robot.
However, we assume the trend lines in Figure \ref{fig:network-trendline}a) are a good approximation of the RTT delay time, therefore we can use the RTT latency of \textit{P.R.L.} and \textit{R.W.P.} as the two extremes for each network, Table \ref{tab:mean-latencies-sim}, to analyse how the \textit{cumulative delay time} is affected.

\begin{figure}[h!]
\begin{center}
\includegraphics[width=0.9\columnwidth]{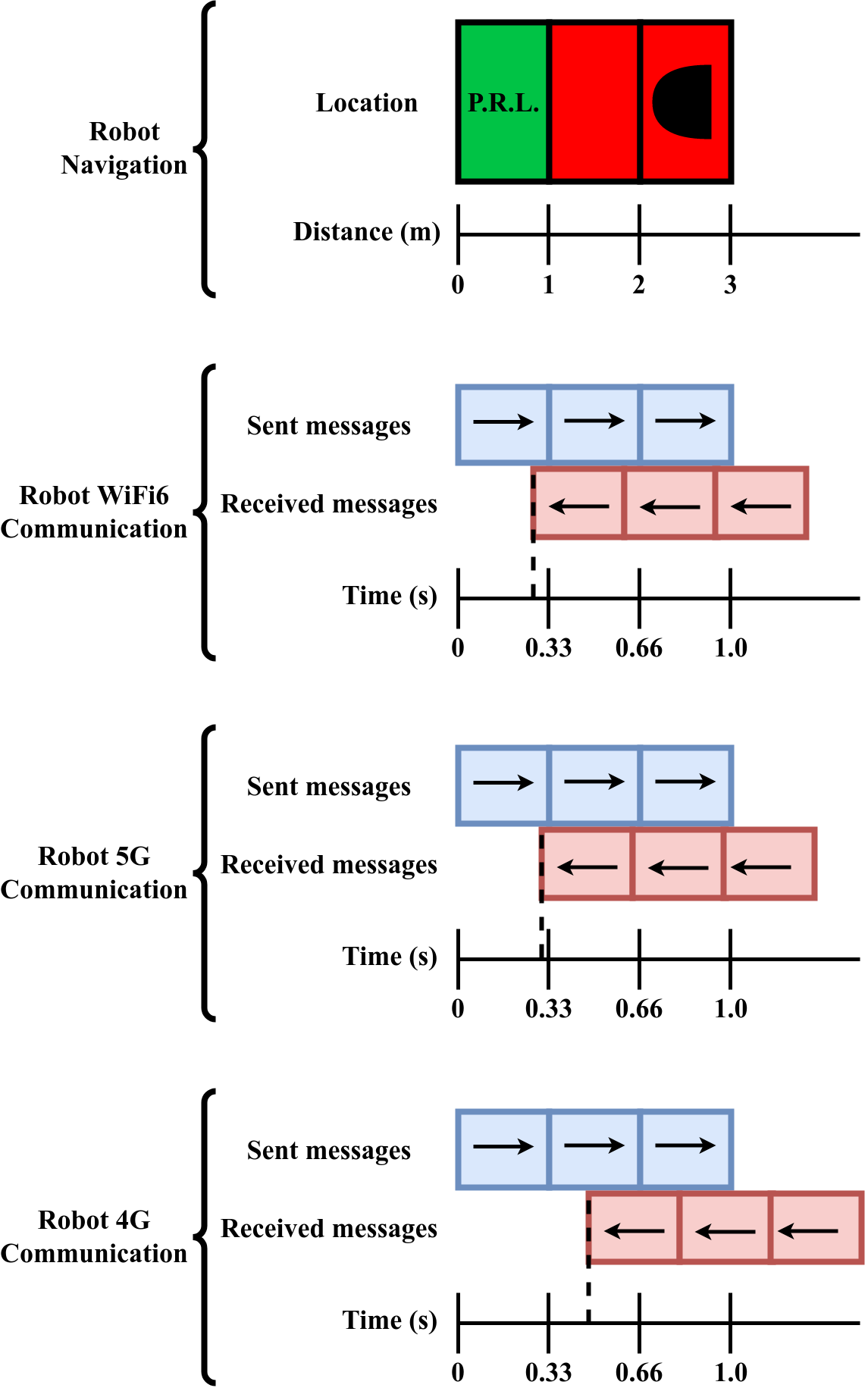}
\caption{Timeline of 1 second, showing the flow of sent location messages and the \textit{cumulative delay time} experienced for data point \textit{P.R.L.} in received commands by the robot using different wireless networks.
The dashed vertical lines show the time of the command message being returned---the elapsed time being the total of:
(i)~the time required to send a location message, 
(ii)~plus the time required to process the message and prepare a command message in response, 
(iii)~plus the time taken to send the command message.
If the vertical dashed line occurs before the next command (blue box) is sent (at time 0.333s), then the localisation will not lag behind.
This is the case for WiFi6 and 5G, but not for 4G.
}
\label{fig:localisation-results-prl}
\end{center}
\end{figure}

\begin{figure}[h!]
\begin{center}
\includegraphics[width=0.9\columnwidth]{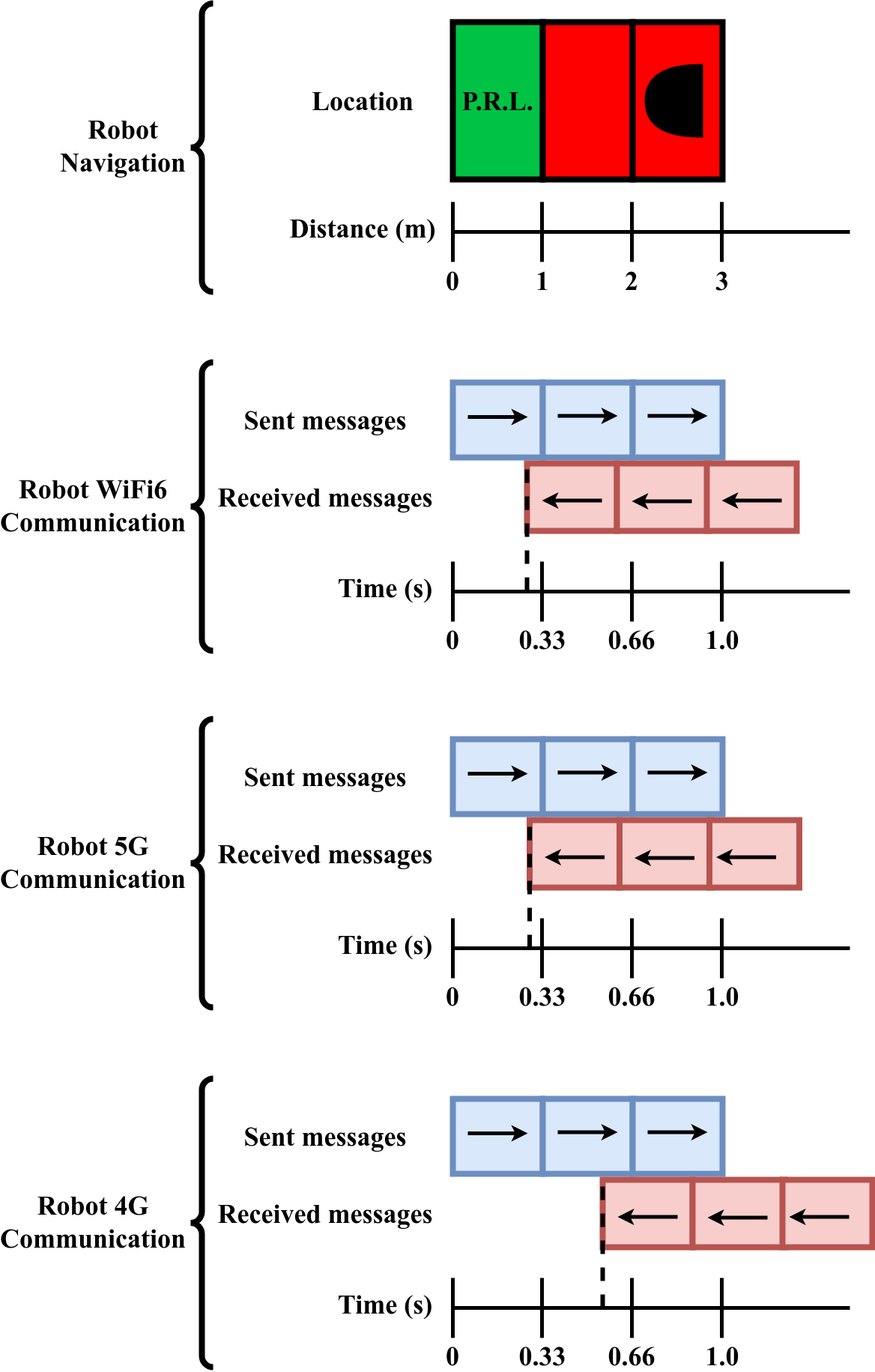}
\caption{Timeline of 1 second, showing the flow of sent location messages and the \textit{cumulative delay time} experienced for data point \textit{R.W.P.} in received commands by the robot using different wireless networks.
The dashed vertical lines show the time of the command message being returned---the elapsed time being the total of:
(i)~the time required to send a location message, 
(ii)~plus the time required to process the message and prepare a command message in response, 
(iii)~plus the time taken to send the command message.
If the vertical dashed line occurs before the next command (blue box) is sent (at time 0.333s), then the localisation will not lag behind.
This is the case for WiFi6 and 5G, but not for 4G.
}
\label{fig:localisation-results-rwp}
\end{center}
\end{figure}

To demonstrate why URLLC is an important feature to 5G it needs to be accurately applied in certain use cases.
As \textit{processing time} is unknown in many use cases it is important and required in the results obtained here, to demonstrate why URLLC can impact localisation and real-time control.
To prove 5G is on track to provide real-time control, even without having URLLC as a feature yet, the robot sends 3 location updates every metre and requires a response within 0.333s (333.3ms) to allow it to carry out an operation while the location has not changed, i.e. in real-time.
Calculating the delta time between the required response time and the cumulative delay time provides lead times for both WiFi and 5G, but lag time for 4G, which is shown in Tables \ref{tab:location-latency-prl} and \ref{tab:location-latency-rwp}, and illustrated by the accompanying Figures \ref{fig:localisation-results-prl} and \ref{fig:localisation-results-rwp}.
This result shows that if 4G was employed for communication, a robot would accumulate an overhead of between 4.7 seconds and 7.0 seconds (due to network lag) in just 10 seconds of travel, and therefore would not be able to operate within real-time.

For further evaluation of the experiment results, we performed simple vertex form quadratic calculations to visualise trend lines and observe the \textit{expected} RTT latency over the 30 metre path of the remote controlled robot, as shown in Figure \ref{fig:network-trendline}.
The evaluation was only performed for 4G and 5G as WiFi6 barely observed any demonstrable change over the 30 metre path.
Moreover, including WiFi6 to Figure \ref{fig:network-trendline}a) greatly reduced the usefulness of the results and made them unclear as it skewed the y-axis in favour of WiFi6.
The trend line for 5G reduces between point \textit{P.R.L.} and \textit{R.W.P.} even though the distance from the access point increases. 
This is because point \textit{R.W.P.} has a more direct and open view of the central access point antenna, which is directly pointing at it and the signal does not have to go over the roof of a nearby building.

\begin{table}[ht]
\begin{tabular}{|c|c|c|c|c|}
\hline
\multicolumn{5}{|c|}{\textit{P.R.L.}} \\ \hline
\multicolumn{1}{|c|}{Network} & 
\multicolumn{1}{c|}{\begin{tabular}[c]{@{}c@{}}Sent \&\\ Rec'vd\\ (ms)\end{tabular}} &
\multicolumn{1}{c|}{\begin{tabular}[c]{@{}c@{}}Proc\\ (ms)\end{tabular}} 
& \multicolumn{1}{c|}{\begin{tabular}[c]{@{}c@{}}Cumu\\ (ms)\end{tabular}} & 
\begin{tabular}[c]{@{}c@{}}Cumu $\Delta$\\ (ms)\end{tabular} \\ \hline
\multicolumn{5}{|p{0.9\columnwidth}|}{Results using Processing delay similar to human reaction time \cite{latency:humanbench}} \\ \hline
\multicolumn{1}{|c|}{WiFi6} & \multicolumn{1}{c|}{1.2} & \multicolumn{1}{c|}{273.0} & \multicolumn{1}{c|}{274.2} & +0.0 (-58.8) \\ \hline
\multicolumn{1}{|c|}{5G} & \multicolumn{1}{c|}{29.5} & \multicolumn{1}{c|}{273.0} & \multicolumn{1}{c|}{302.5} & +0.0 (-30.5) \\ \hline
\multicolumn{1}{|c|}{4G} & \multicolumn{1}{c|}{216.7} & \multicolumn{1}{c|}{273.0} & \multicolumn{1}{c|}{489.7} & +4701.0 (+156.7) \\ \hline
\label{tab:location-latency-prl}
\end{tabular}
\caption{\textit{Cumulative Delay Time} latency at each step of message transmission and overall \textit{cumulative $\Delta$ time}, showing individual message processing time (proc), cumulative (cumu) delay time and cumulative lead/lag time difference, or $\Delta$, in milliseconds (ms).}
\end{table}

\begin{table}[ht]
\begin{tabular}{|c|c|c|c|c|}
\hline
\multicolumn{5}{|c|}{\textit{R.W.P.}} \\ \hline
\multicolumn{1}{|c|}{Network} &
\multicolumn{1}{c|}{\begin{tabular}[c]{@{}c@{}}Sent \&\\ Rec'vd\\ (ms)\end{tabular}} &
\multicolumn{1}{c|}{\begin{tabular}[c]{@{}c@{}}Proc\\ (ms)\end{tabular}} & 
\multicolumn{1}{c|}{\begin{tabular}[c]{@{}c@{}}Cumu\\ (ms)\end{tabular}} &
\begin{tabular}[c]{@{}c@{}}Cumu $\Delta$\\ (ms)\end{tabular} \\ \hline
\multicolumn{5}{|p{0.9\columnwidth}|}{Results using Processing delay similar to human reaction time \cite{latency:humanbench}} \\ \hline
\multicolumn{1}{|c|}{WiFi6} & \multicolumn{1}{c|}{1.3} & \multicolumn{1}{c|}{273.0} & \multicolumn{1}{c|}{274.3} & +0.0 (-58.7) \\ \hline
\multicolumn{1}{|c|}{5G} & \multicolumn{1}{c|}{22.9} & \multicolumn{1}{c|}{273.0} & \multicolumn{1}{c|}{295.9} & +0.0 (-37.1) \\ \hline
\multicolumn{1}{|c|}{4G} & \multicolumn{1}{c|}{294.0} & \multicolumn{1}{c|}{273.0} & \multicolumn{1}{c|}{567.0} & +7020.0 (+234.0) \\ \hline
\label{tab:location-latency-rwp}
\end{tabular}
\caption{\textit{Cumulative Delay Time} latency at each step of message transmission and overall \textit{cumulative $\Delta$ time}, showing individual message processing time (proc), cumulative (cumu) delay time and cumulative lead/lag time difference, or $\Delta$, in milliseconds (ms).}
\end{table}

\subsection{Discussion}
\label{sec:exp2:discussion}
The distance to point \textit{P.R.L.} is 49.1 m and mean RTT latency of 22.9 ms and \textit{R.W.P} is 72.0 m and mean RTT latency of 63.9 ms, for the 5G network, which means that one-way communication latency is 11.5 ms and 32.0 ms, respectively.
The latency is low enough to enable live 30FPS video streaming, but not low enough to allow for real-time tracking and control.

\section{Conclusion}
\label{sec:conclusion}

This work draws two important conclusions. 
Firstly, it evaluates the performance of a private 5G-SA telecommunications network, a private WiFi6 network and public 4G telecommunications network for the use case of high throughput and low latency operations.
Experiments in Section \ref{sec:physical:exp} were conducted in the context of an agricultural use case: a robot capturing images in a field, streaming that video to an off-board edge computer for identifying weeds and sending actuation commands back to the robot or to a human user on another computer.
The results demonstrated that public 4G cannot be used in agriculture to support high throughput and low latency operation.
Further, in our controlled setting, we found that WiFi6 performed better than 5G.
WiFi6 never saturated during throughput testing, whilst 5G saturated at approximately 60Mbps when testing 1-RGBD video streaming.
According to~\citep{ookla:5g}, the achieved throughput is higher than leading countries' public 5G results from gathered data in Q1-Q2 of 2021.
However, these results show a good outcome overall for 5G as it shows that the technology is still maturing.
WiFi6 had a lower latency on average of 18.2ms compared to that of 5G.
The 5G mast is further by 45.4 metres and 114.0 metres in the Vegetable Polytunnel and Walled Garden, respectively, compared to the WiFi6 router.
The greater distance from the access point further contributes to the worse performance in the Walled Garden for the 5G network.
However, this highlights the 5G network's coverage over a greater distance and, a feature not tested, support for connecting a greater amount of devices.
WiFi/WiFi6 routers can support a few devices, any increase in number of devices can greatly increase complexity.
Whereas, the 5G network can inherently support a greater number of devices with gradual increase in complexity. 
It is worth noting that the obtained results are only a snapshot of the private 5G performance at the time of data collection. 
The 5G network is continuously being updated and improved, making it more robust and balancing the upload and download ratio for different use cases.

Secondly, simulation experiments were conducted, in order to assess the viability of performing a more complex hypothetical variant of our agricultural use case using each of the three network setups.
Specifically, these experiments analysed latency.
As previously observed, these results reaffirmed that 4G is too slow to be able to perform the task at hand.
The WiFi6 and 5G produced sufficient speed to manage the job.
Furthermore, the results showed that only in extreme cases, where the processing time is longer or the velocity of the robot is greater, will WiFi6 have advantage over 5G.

In conclusion, the results in this body of work are significant for the agricultural domain. 
They clearly identify strengths and weaknesses of current and state-of-the-art wireless network infrastructures in rural environments.
Moreover, the results identify the fundamental requirements that the future smart farm will have for the telecommunications industry.
It is clear that 4G cannot support agricultural activities, and the lower coverage, higher attenuation and much slower commercial uptake of WiFi6 make it an impractical solution.
Finally, this work highlights that there is no single wireless network that is best suited for agri-technology and agri-robotics, but using a mixture of the state-of-the-art can provide a better solution.
For example, private 5G can be used to move data faster between longer distances connected to a WiFi6 (or multi-WiFi) wireless backhaul that extends to locally connected robots and sensors in a farm field.

The next steps with this line of research involve testing more complex scenarios in a physical environment.
This includes the hypothetical setup simulated in Section~\ref{sec:sim:exp}, as well as setups with multiple robots in the field, larger fields (where the distance to the network mast is greater) and more complex actuation messages going to the robot such that send and receive transmissions are more balanced than in the experiments presented here.
As public 5G roll-out continues world-wide, having better understanding of the benefits in agriculture will help farmers make the case for rural deployments of such networks.
The contribution of the work shared here helps to demonstrate that the wireless infrastructure of 5G is required to facilitate even the most basic precision agriculture use case.


\clearpage
\appendix
\section{Appendix}
\label{appendix}
\begin{figure*}
  \includegraphics[width=\textwidth,height=18cm]{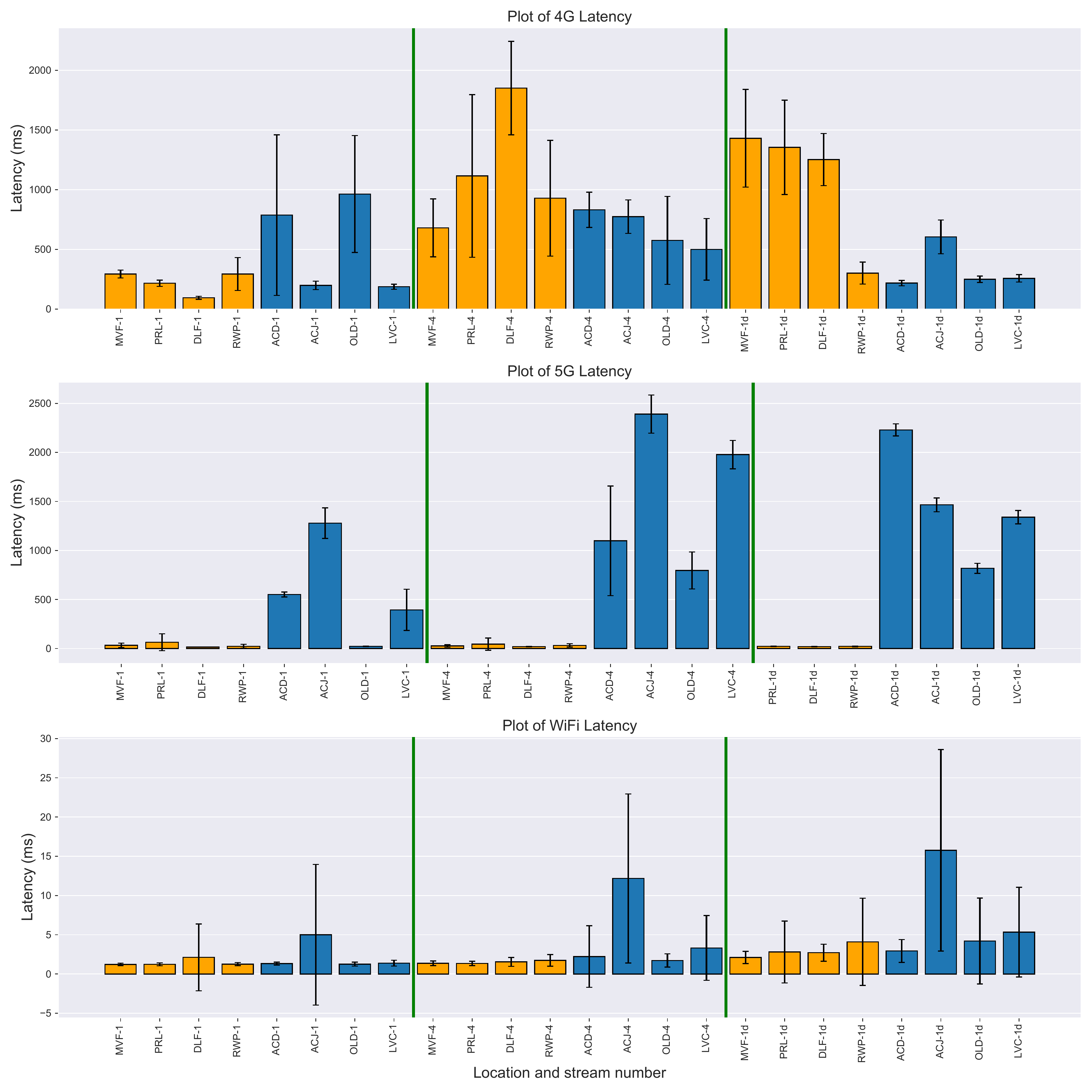}
  \caption{Latency results across all performance metrics and parameters. The vertical lines separate the data stream type and the orange coloured bars represent the Vegetable Polytunnel and the blue coloured bars represent the Walled Garden.}
  \label{fig:latency-combined}
\end{figure*}

\begin{figure*}
  \includegraphics[width=\textwidth,height=18cm]{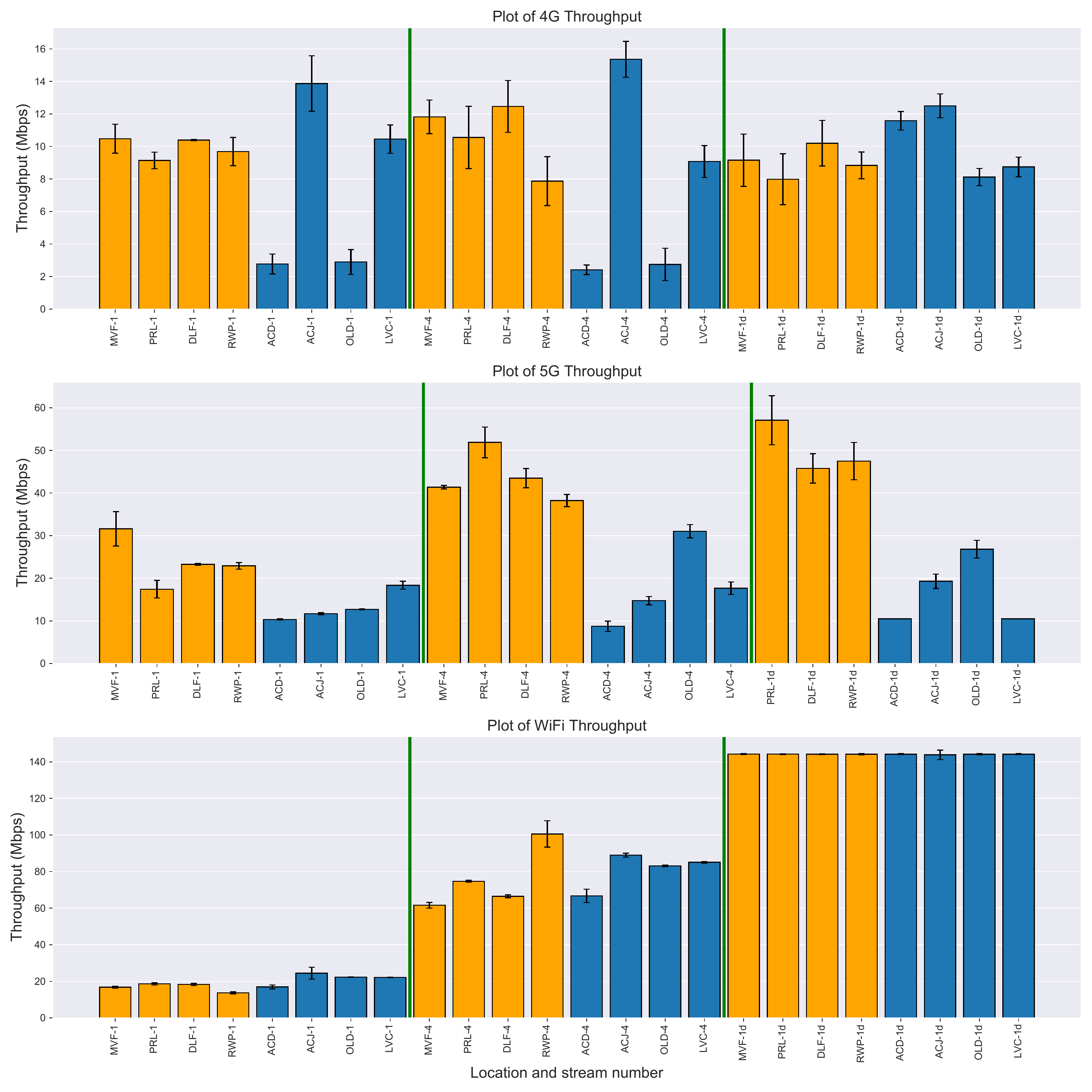}
  \caption{Throughtput results across all performance metrics and parameters. The vertical lines separate the data stream type and the orange coloured bars represent the Vegetable Polytunnel and the blue coloured bars represent the Walled Garden.}
  \label{fig:throughput-combined}
\end{figure*}

To give a complete visual picture of our findings and data collection from experiments in Section~\ref{sec:physical:exp}, we have collated and plotted all the data in simple and easy to read graphs.
The data is split into two main figures, each figure represents one of the two main performance metrics being analysed, i.e. latency in Figure~\ref{fig:latency-combined} and network throughput in Figure~\ref{fig:throughput-combined}.
Each figure contains 3 subplots and each subplot represents a wireless network, i.e. 4G, 5G, WiFi6.
Finally, each subplot is split by a vertical line into three sections, highlighting the data stream network parameter, and bar colour represents one of the two locations where experiments were conducted.

The public 4G latency performance in Figure \ref{fig:latency-combined} is poor throughout all streaming experiments and in all environments.
Unlike WiFi6 and the private 5G, for 4G it is difficult to analyse if the environment or the different streaming experiments cause an increase in latency, this is because the RF interference over a larger distance is impossible to predict.
However, it can be confirmed that the latency is far too high for real-time video streaming, regardless of what type of streaming experiment is conducted.

The latency for 5G in Figure \ref{fig:latency-combined} is extremely low, and it is close to WiFi6 in the Vegetable Polytunnel environment (orange coloured bars).
However, in a distant environment, obstructed by tree cover and a wall, it suffers greatly and in certain parts of the environment the latency is as bad or worse than the public 4G (ACJ-4).

The latency results in Figure \ref{fig:latency-combined} for WiFi6 standard deviation indicates negative latency, which is impossible. 
This is because the latency is so low and on occasion it can spike making the negative portion of the standard deviation dip below zero. 
This makes WiFi6's standard deviation negligible, it is kept for illustrative purposes.
The main increase in latency for WiFi6 can be seen during the RGB-D data streaming experiments and when operating in an open field in the Walled Garden.
This is expected for WiFi6 as signal loss in an open field is far greater than in an indoor space or a space with many walls and obstacles.
The latency still remains extremely low.

The public 4G throughput results in Figure \ref{fig:throughput-combined} are interesting, as regardless of streaming experiment they hit a certain limit of throughput.
As suggested, from our own experiments on public 4G and from the data obtained from \cite{Ofcom}, the maximum and mean throughput (upload speed) should be between 20Mbps and 14.7Mbps, which is what we see. Albeit, there are some experiment locations that have much lower throughput, which could be caused by many factors, e.g., RF interference, increased traffic load, traffic load optimisation, etc. 
Therefore, we can assume that we are saturating the upload speed of the public 4G network and we cannot expect much higher throughput.

The 5G throughput results are impressive, and clearly much higher than 4G.
However, if we examine the throughtput results between WiFi6 and 5G, specifically for the RGB-D streaming experiment, we can see that the 5G network has also saturated in terms of upload speed.
We can assume that the 5G network maximum upload speed is close to 65Mbps.
It was never the intention of this body of work to find the maximum upload speed of the particular configuration of the 5G network setup at the University of Lincoln.
Because, the 5G network is continuously being improved, and for example UL/DL ration in the future can be configurable.
For the current release of 5G-SA N77 it is not (at least not to our knowledge).
Moreover, there are different 5G network technologies and different iterations of 5G that will perform completely differently to each other, we would not be contributing to the field by specifically finding the limits of our particular system, which itself is continually evolving.

The WiFi6 throughput results are almost perfectly aligned with theoretical expectations.
The RGB data stream is compressed and throughput increases only if movement is detected and there are many different colour changes in very fast succession in front of the camera, which does not occur in our green and brown images, the throughput is variable and unpredictable.
However, for the RGB-D experiments, the data stream is still compressed, but at a static rate.
This means that the data streamed should always be the exact same regardless of how fast the scene in front of the camera changes and regardless of colour changes.
Theoretically, this value should be 144Mbps (or 18MBps), which is what WiFi6 approximately reaches during the RGB-D experiments.
Clearly, WiFi6 can stream the data it is expected to, and we have no reached a saturation limit of upload or download.
However, the latency results, which are excellent show the one weakness of WiFi6.
In an outdoor open field environment the signal loss will be exponential.

\printcredits
\clearpage

\bibliographystyle{model1-num-names}
\bibliography{mainbib}





\end{document}